# Introductory E & M Lab Manual for Virtual Teaching

*Neel Haldolaarachchige
Department of Physical Science, Bergen Community College, Paramus, NJ 07652

Kalani Hettiarachchilage
Department of Physics, Seton Hall University, South Orange, NJ 07962

December 24, 2020

*Corresponding Author: haldo.physics@gmail.com

**ABSTRACT**

Introductory electricity and magnetism lab manual was designed to use with virtual Physics II class. The lab manual consists of experiments on electrostatics, electric potential and energy, current and resistance, DC circuits, electromagnetism and AC circuits. Virtual experiments were based on simulations. Open educational resources (OER) were used for all experiments. Virtual experiments were designed to simulate in person physical lab experiments. Special emphasis was given to computational data analysis with excel. Formatted excel sheets per each lab were given to students and step by step calculation in excel were explained during the synchronous class. Learning management system (LMS) was used to fully web enhance the lab class. Virtual labs were delivered by using live video conference technology and recorded lab sessions were added to LMS. Lab class were tested with both virtual delivery methods (synchronous and asynchronous). Student learning outcomes (understand, apply, analyze and evaluate) were studied with detailed lab reports and end of the semester lab based written exam which confirmed the virtual lab class was as effective as the in person physical lab class.

**CONTENTS**

# EXPERIMENT 1 ELECTROSTATIC FORCE, FIELD AND EQUIPOTENTIAL-LINES

## OBJECTIVE

Electric field generated around electric charges is investigated on a few different shapes of electrodes. Electric field maps are produced by using electric potential measurements. Electric field magnitudes between electrodes are calculated. Electric charge magnitude is calculated for points like electrodes.

## THEORY AND PHYSICAL PRINCIPLES

In nature there are two types of charges which are called negative and positive charges. These charges originate at the atomic level and negative charge is due to charge of electrons and positive charge is due to charge of the proton.

There is a force between two charge particles which is called the Coulomb force. It is an attractive force if the particles are oppositely charged (positive and negative) and a repulsive force if the charges of particles are the same (positive/positive or negative/negative).

Electrostatic force between two charge,

$$\vec{F} = \frac{1}{4\pi\varepsilon_0}\frac{q_1 q_2}{r^2}\hat{r} = \frac{k q_1 q_2}{r^2}\hat{r} \quad (1)$$

k is called the Coulomb constant and $\varepsilon_0$ is called the permittivity of free space.

$$k = 8.98 \times 10^9 \frac{Nm^2}{C^2} \quad (2)$$

$$\varepsilon_0 = \frac{1}{4\pi k} = 8.85 \times 10^{-12} \frac{C^2}{Nm^2} \quad (3)$$

The electric potential, V, can be computed by dividing the distance from the charge, r, into the product of the charge's magnitude, Q, and coulomb constant k.

$$V = \left(\frac{kQ}{r}\right) \quad (4)$$

$$Q = \frac{Vr}{k} \quad (5)$$

Magnitude of the charge on the electrodes can be calculated by using the voltage between electrodes and separation between equipotential lines.

The electric field, E, can be computed by dividing the change in distance, delta d, into the change in the electric potential, delta V.

$$E = \left(\frac{\Delta V}{d}\right) \quad (6)$$

The electric field, E, can be computed by dividing the test charge relatively close to the main charge Q , q, into the force being exercised on the test charge, F.

$$E = \frac{F}{q} \quad (7)$$

The electric field, E, can be computed by dividing the distance from point a and b, $d_{ab}$, into the potential difference from point a and b, $V_a$ and $V_b$.

$$E = \left(\frac{V_a - V_b}{d_{ab}}\right) \quad (8)$$

*Equipotential lines*

When the same electric potential points around a charged particle are connected to each other it will create an equipotential line which is a contour map around the charge particle. Equipotential lines are perpendicular to electric field lines.

Contour map of electric potential around a charged object depends on the shape of the object.

Electric field strength between equipotential lines can be calculated by knowing potential values of each equipotential line and the distance between them.

## APPARATUS AND PROCEDURE

*Part A: Investigation of coulomb force between electric charges*

- Part A of the experiment is done with following simulation:
- https://phet.colorado.edu/en/simulation/coulombs-law
- A very detail video lesson of virtual lab (data collection with simulator and data analysis with excel) can be found here: https://youtu.be/Y39A8JZJHDE

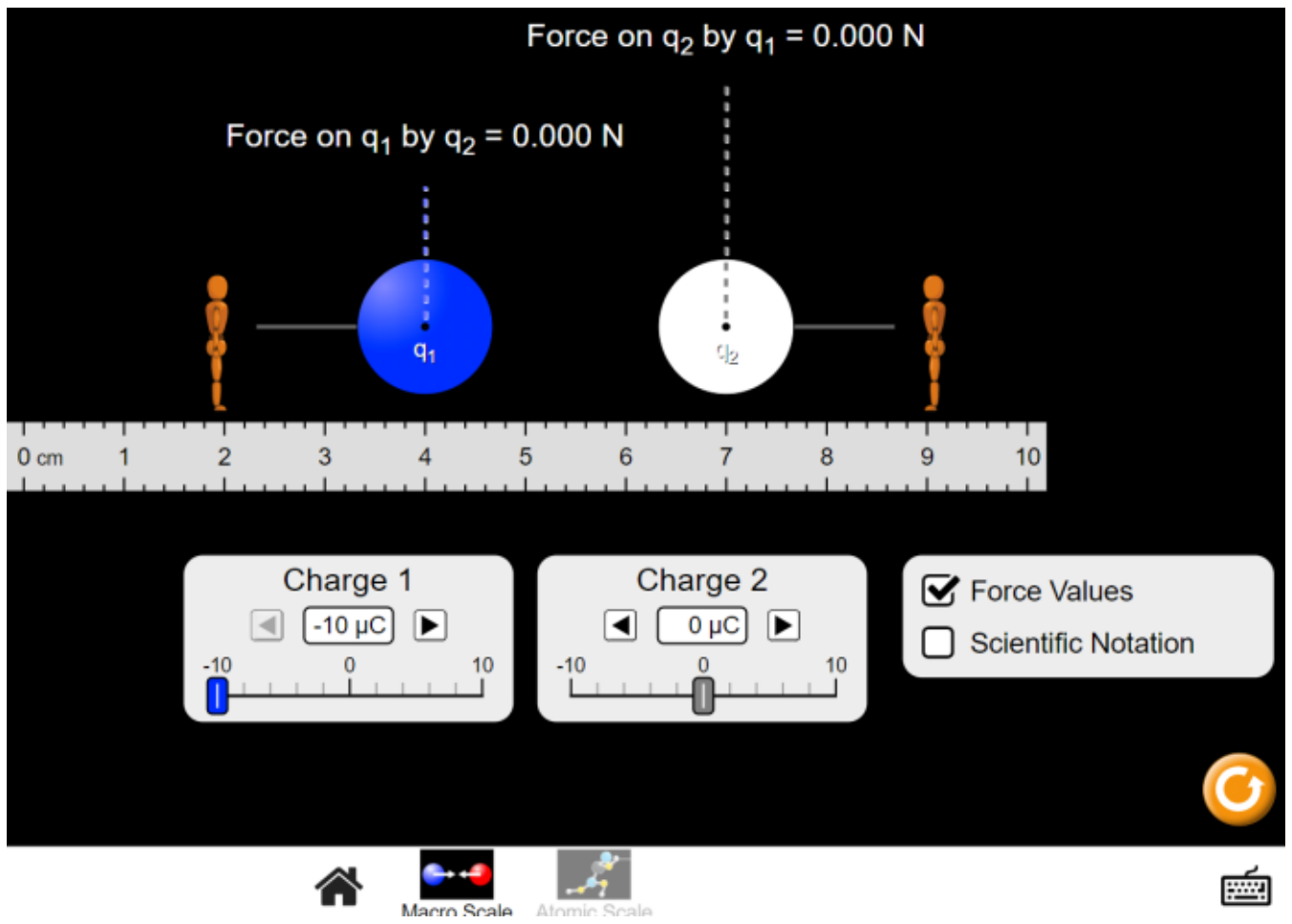

Figure 1 Electrostatic force simulation (Picture credit: https://phet.colorado.edu)

- Set the charge of the object-1 to +10.0 μC and place it at 0.0cm location.
- Set the charge of the object-2 to +10.0 μC and place it as close as possible to object-1.
- Measure electrostatic force acting on charge objects.
- Calculate the electrostatic force on objects by using Coulomb's law.
- Compare observed and calculated coulomb force between charged objects by calculating percent difference.
- Then repeat the above procedure by changing the value of charged object-2 in steps of 1.0 μC at a time.
- Keep the charge of both objects to 10.0 micro Coulombs.
- Then, move the object-2 1.0cm at a time away from the object-1 and calculate the coulomb force for each case.
- Calculate the electrostatic force on objects by using Coulomb's law.
- Compare observed and calculated coulomb force between charged objects by calculating percent difference.
- Then make a graph of force vs charges and explain the behavior in terms of Coulomb force.
- Make a graph of force vs distance and explain the behavior in terms of Coulomb force.

*Part B: Electric potential map and electric field lines.*

- Part B of the experiment is done with following simulation:
- https://phet.colorado.edu/en/simulation/charges-and-fields

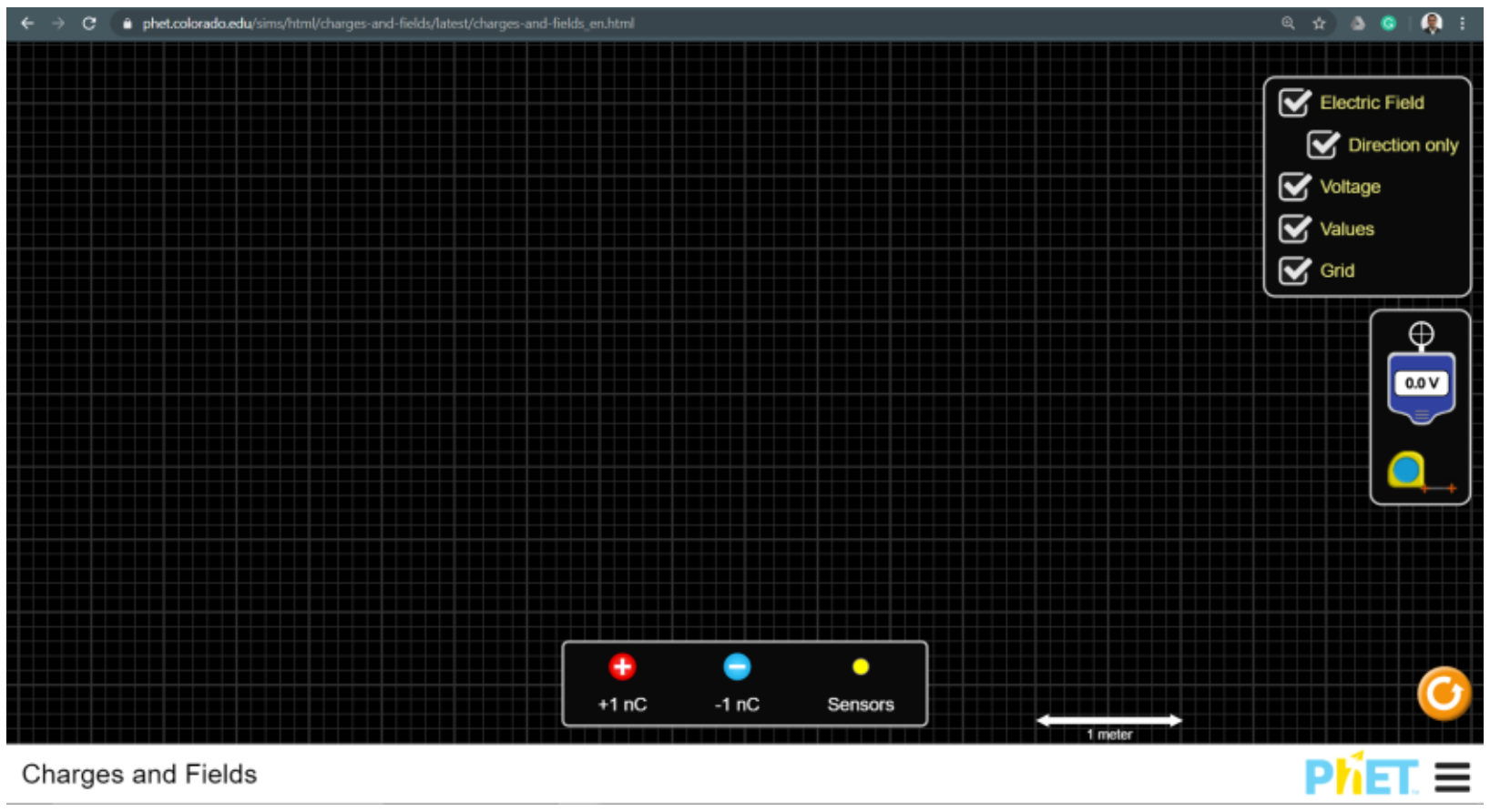

Figure 2 Simulation of equipotential lines and electric field map (Picture credit: https://phet.colorado.edu)

- Place one positive charge on the map and draw equipotential lines after each 50.0cm from the charge.
- Measure distance and potential and complete the table-3.
- Set the two-point charges (one positive and one negative) on the grid and separate them about eight large squares in the simulator grid.
- Then draw equipotential lines in between point charges (one equipotential line per every large square line).
- Switch on electric field lines and save a picture of dipole charge electric field and equipotential map.
- Then make positive plate charge by combining point-like charges and use points like negative charge. Then draw equipotential lines in between point charges (one equipotential line per every large square line). Switch on electric field lines and save a picture of electric fields and equipotential maps of new electrodes.
- Then make positive and negative plate charges by combining points-like charges. Then draw equipotential lines in between point charges (one equipotential line per every large square line). Switch on electric field lines and save pictures of electric fields and equipotential maps of new electrodes.

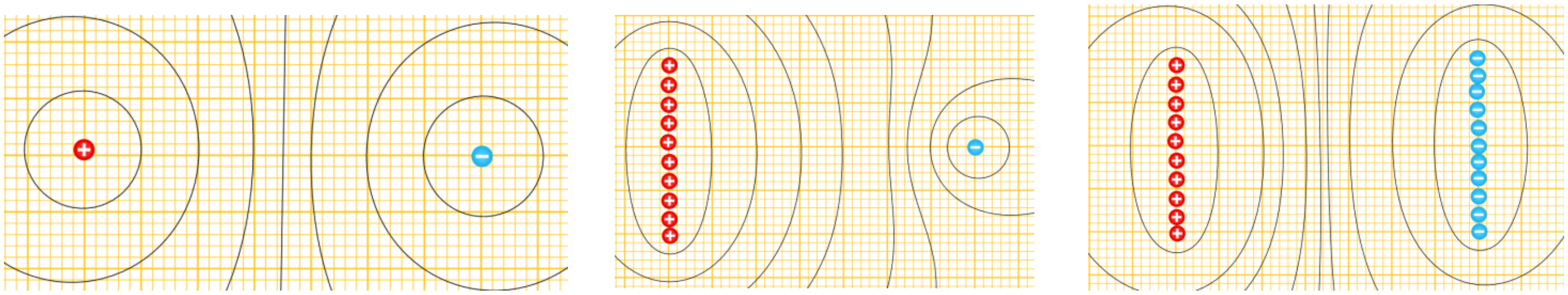

(a) (b) (c)

Figure 3 Equipotential maps for different shape of electrodes, a) point charge, b) plate change and a plate, c) two plates (Picture credit: https://phet.colorado.edu)

## PRE LAB QUESTIONS

1) Describe the electric field lines?
2) Where does the electric field line start and end?
3) Describe the equipotential lines?
4) Where is the equipotential line starts and ends?
5) Describe work that needs to be done to move a charge particle between nearby equipotential lines?
6) Describe work that needs to be done to move a charge particle on an equipotential line?

## POST LAB QUESTIONS

*Electric field hockey with simulation.*

- This should be done by using following simulation:
- https://phet.colorado.edu/sims/cheerpj/electric-hockey/latest/electric-hockey.html

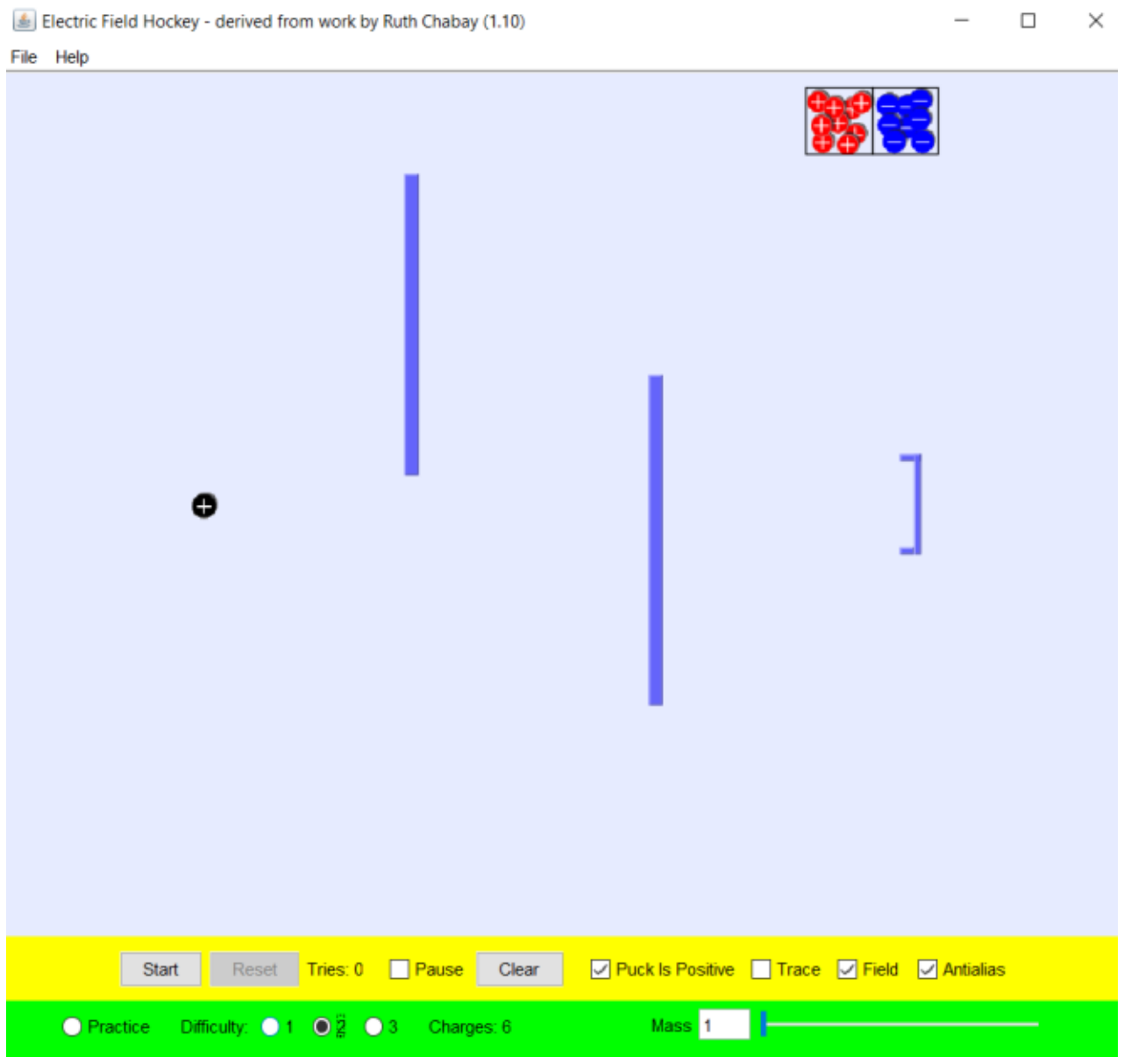

Figure 4 Electric hockey simulation (Photo credit: https://phet.colorado.edu)

- This simulation works directly in a web browser.
- Black positive charge in middle (left side) should be sent into the goal (blue rectangular bracket in the right middle).
- Blue straight lines are barriers which means black charge must go to the goal without colliding barriers.
- Set the difficulty level to 1 and click on trace.
- You should use positive/negative charges (in buckets in top right) in different places to put the black charge towards the goal.
- To test your setup just click the start button and see the path of the balck charge.
- Rearrange the other (positive/negative) charges and try again.
- Repeat the procedure till you get the goal.
- After you get the goal take a screenshot with the trace is shown and attach it to your lab report.
- You must do this at least for two difficulty levels.
- Attached screenshots of electric field hockey for difficulty level 1 and 2.
- Discuss about the electric repulsion and attraction force and the electric field in the case of electric hockey.

## DATA ANALYSIS AND CALCULATIONS

*Part A: Investigation of Coulomb's force*

Table 1 Electrostatic force analysis as a function of charges

| Charge-1 $Q_1$ [ ] | Charge-2 $Q_2$ [ ] | $Q_1Q_2$ [ ] | Separation R [ ] | Force observed F_obs [ ] | Force calculated F_cal [ ] | Percent difference [ ] |
|---|---|---|---|---|---|---|
| | | | | | | |
| | | | | | | |
| | | | | | | |
| | | | | | | |
| | | | | | | |
| | | | | | | |
| | | | | | | |
| | | | | | | |
| | | | | | | |
| | | | | | | |

Table 2 Electrostatic force analysis as a function of separation

| Charge-1 $Q_1$ [ ] | Charge-2 $Q_2$ [ ] | Separation R [ ] | $R^2$ [ ] | Force observed F_obs [ ] | Force calculated F_cal [ ] | Percent difference [ ] |
|---|---|---|---|---|---|---|
| | | | | | | |
| | | | | | | |
| | | | | | | |
| | | | | | | |
| | | | | | | |
| | | | | | | |
| | | | | | | |
| | | | | | | |
| | | | | | | |
| | | | | | | |

- Make a graph of F_obs vs $Q_1Q_2$ for table-1. Then discuss the behavior of the graph in terms of coulomb's law.
- Make a graph of F_obs vs $R^2$ for table-1. Then discuss the behavior of the graph in terms of coulomb's law.

*Part C: Equipotential and electric field lines*

- Attached all three graphs that you created for different types of electrodes.
- Discuss the shapes of equipotential lines or each of the graphs.
- Discuss the shape of the electric field lines for each of the graphs.
- Complete the following table to calculate the electric field for graph-1 and 2.

Table 3 Electrostatic field and charge of particle of point like electrodes

| Potential [ ] | Radius [ ] | Electric field for point like electrodes [ ] | Charge calculated [ ] | Percent error [ ] |
|---|---|---|---|---|
| | | | | |
| | | | | |
| | | | | |
| | | | | |
| | | | | |

Table 4 Electric field analysis of other electrodes

| Plate like electrodes | | | Point and plate electrodes | | |
|---|---|---|---|---|---|
| Potential [ ] | Radius [ ] | Electric field [ ] | Potential [ ] | Radius [ ] | Electric field [ ] |
| | | | | | |
| | | | | | |
| | | | | | |
| | | | | | |
| | | | | | |

# EXPERIMENT 2 OHM'S LAW AND RESISTIVITY

## OBJECTIVE

Ohm's law is investigated by using simple direct current (DC) circuits. Current and voltage behavior across a resistor will be investigated and linear behavior of them is used to confirm the Ohm's law. Non ohmic behavior is investigated by using a graph of voltage vs current and resistivity of a wire is investigated by using Ohm's law.

## THEORY AND PHYSICAL PRINCIPLES

Ohm's law is one of the simplest yet very useful laws of electricity and magnetism. This law explains the linear behavior of current and voltage across certain materials which are called the Ohmic type materials. Most of the general use electronic instruments contain many Ohmic type resistors. On the other hand, material not following Ohm's law is called non Ohmic type.

Ohm's law: $V \propto I \ \rightarrow V = IR$ (1)

V is voltage, I is current, and R is resistance.

$$R = \frac{V}{I} \rightarrow \frac{volt\ (V)}{amp\ (A)} = \text{Ohm}\ (\Omega) \tag{2}$$

Resistance (R) is an intrinsic property of an object and it depends on the shape of the object and materials of the object.

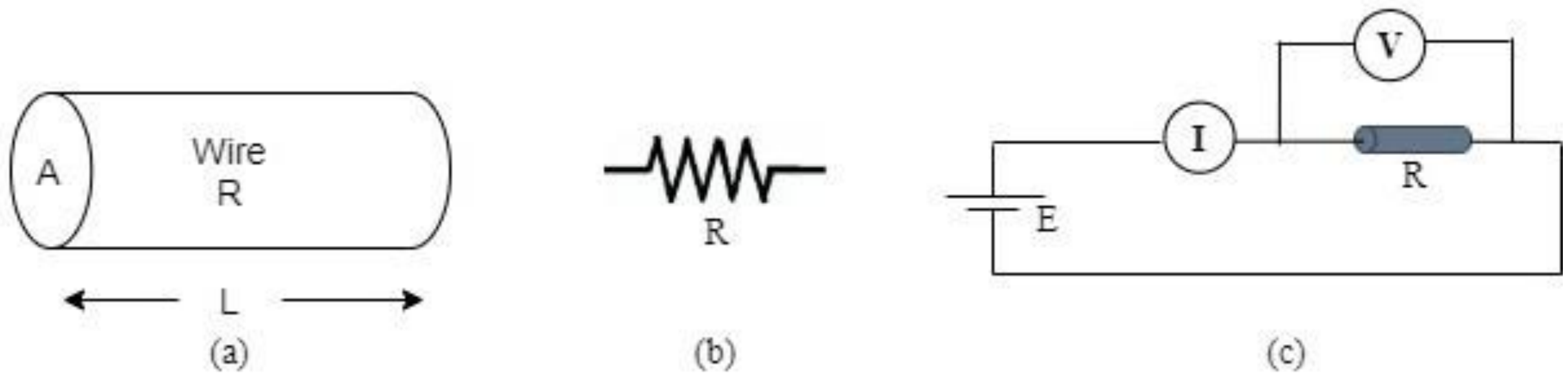

Figure 1 (a) Geometric characteristics of wire (conductor), (b) circuit symbol of a resistor, (c) simple circuit with resistor connected to battery, ammeter and voltmeter commented to the circuit

Resistance of the wire is directly proportional to length and inversely proportional to cross section area. And the proportionality constant is called the resistivity of the material which the wire made of.

$$R \propto \frac{L}{A} \rightarrow R = \rho \frac{L}{A} \tag{3}$$

$$A = \pi r^2 \tag{4}$$

$$\rho = \frac{RA}{L} \tag{5}$$

$$V = \frac{\rho I}{A} L \tag{6}$$

Resistivity of the wire can be investigated experimentally by measuring voltage across the wire as function of length of the wire.

## EQUIPMENT AND PROCEDURE

- This experiment is done with simulation and click here: https://phet.colorado.edu/en/simulation/circuit-construction-kit-dc-virtual-lab
- A very detail video lesson of virtual lab (data collection with simulator and data analysis with excel) can be found here: https://youtu.be/mIHve8OfkXY and https://youtu.be/ecPEgrMG67s

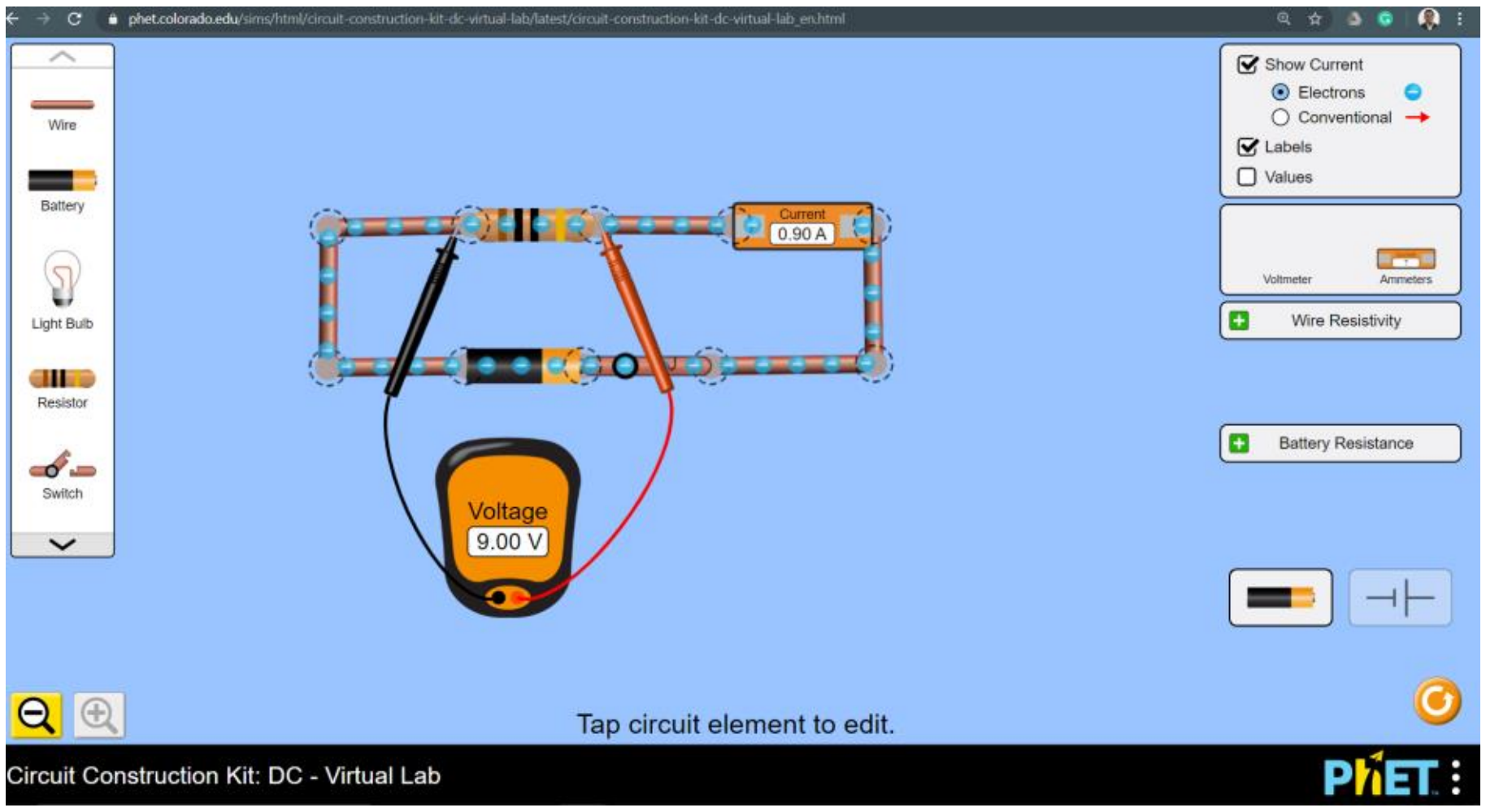

Figure 2 Simple DC circuit simulation (Picture credit: https://phet.colorado.edu)

- Setup a simple circuit with resistor, battery and switch.
- Connect ammeter to the circuit serially to measure current through the circuit.
- Connect voltmeter across resistor to measure the voltage across the resistor.
- Increase voltage of the battery 1.0V at a time and measure current and voltage across the resistor.
- Repeat last procedure for two other different resistor values.
- Replace the resistor with a bulb (consider this is the unknown resistance).
- Measure voltage and current across the bulb by increasing voltage of the battery 1.00V at a time.
- Repeat measurement for two different resistor values of bulb.
- Resistivity of the wire can be done with the following simulation: http://amrita.olabs.edu.in/?sub=1&brch=6&sim=22&cnt=4

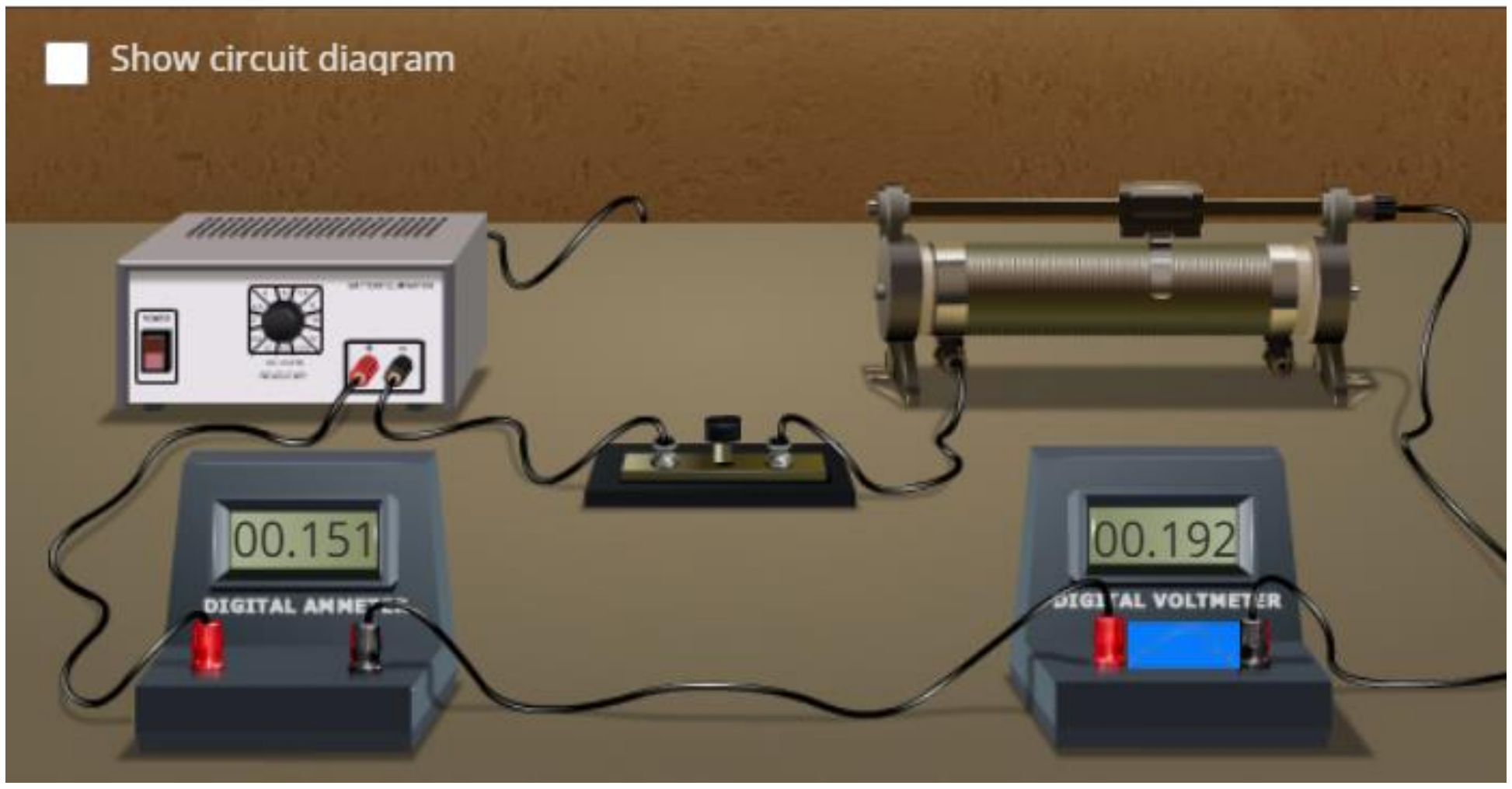

Figure 3 Simple DC circuit simulation (Picture credit: http://amrita.olabs.edu.in/)

- Set the metal, wire length, diameter by using the selection tools in the left side of the simulator.
- Set the resistance of the rheostat to the highest possible value.
- Circuit can be made by clicking and dragging the mouse from one connecting terminal to the other connecting terminal of the devices to be connected.
- Voltage across and the current through the wire can be changed by moving rheostat contact.

**PRE LAB QUESTIONS**

1) Describe Ohm's law?
2) Describe the difference between resistance and resistivity?
3) Describe Ohmic type materials?
4) Describe non-Ohmic type materials?
5) Describe semiconductor and insulator?

**POST LAB QUESTIONS**

1) Describe the behavior of hand and a dog by connecting and checking them with the simple DC circuit?
2) Describe the behavior of eraser and dollar-bill by connecting and checking them with the simple DC circuit?
3) Describe the use of rheostat in the second circuit?

## DATA ANALYSIS AND CALCULATIONS

*Part I: Ohm's law*

Table 1 Voltage and current measurements for selected resistances

| Resistor 1 ($R_1$)<br>$R_1$ = | | | Resistor 2 ($R_2$)<br>$R_2$ = | | | Resistor 3 ($R_2$)<br>$R_3$ = | | |
|---|---|---|---|---|---|---|---|---|
| V<br>[ ] | I<br>[ ] | $R_1^{cal} = \frac{V}{I}$<br>[ ] | V<br>[ ] | I<br>[ ] | $R_2^{cal} = \frac{V}{I}$<br>[ ] | V<br>[ ] | I<br>[ ] | $R_3^{cal} = \frac{V}{I}$<br>[ ] |
| | | | | | | | | |
| | | | | | | | | |
| | | | | | | | | |
| | | | | | | | | |
| | | | | | | | | |
| | | | | | | | | |
| | | | | | | | | |
| | | | | | | | | |
| | | | | | | | | |

Table 2 Resistance and percentage error calculations

| Resistor | Calculated average<br>[ ] | From Graph<br>[ ] | Percent Error between R(known) and R(calculated average) | Percent error between R(known) and R(from graph) |
|---|---|---|---|---|
| R1 | | | | |
| R2 | | | | |
| R3 | | | | |

*Part II: Finding unknown resistance by using Ohm's law*

Table 3 Voltage and current measurements for unknown resistances

| Unknown Resistance 1 ($R_1$) | | | Unknown Resistance 2 ($R_2$) | | |
|---|---|---|---|---|---|
| V [ ] | I [ ] | $R_1^{cal} = \frac{V}{I}$ [ ] | V [ ] | I [ ] | $R_2^{cal} = \frac{V}{I}$ [ ] |
| | | | | | |
| | | | | | |
| | | | | | |
| | | | | | |
| | | | | | |
| | | | | | |
| | | | | | |
| | | | | | |
| | | | | | |

Table 4 Resistance and percentage error calculations

| Resistor | Calculated average [ ] | From Graph [ ] | Percent Error between R(known) and R(calculated average) | Percent error between R(known) and R(from graph) |
|---|---|---|---|---|
| $R_1$ | | | | |
| $R_2$ | | | | |

*Part III: Resistivity of a wire*

Table 5 Voltage and current measurements for resistivity of a wire

| Wire-1 | | Wire-2 | | Wire-3 | |
|---|---|---|---|---|---|
| Material =<br>Length =<br>Diameter = | | Material =<br>Length =<br>Diameter = | | Material =<br>Length =<br>Diameter = | |
| $V$<br>[ ] | $I$<br>[ ] | $V$<br>[ ] | $I$<br>[ ] | $V$<br>[ ] | $I$<br>[ ] |
| | | | | | |
| | | | | | |
| | | | | | |
| | | | | | |
| | | | | | |
| | | | | | |
| | | | | | |
| | | | | | |
| | | | | | |

Table 6 Resistivity and percent error calculations

| Resistivity | Known<br>[ ] | From Graph<br>[ ] | Percent error between ρ(known) and ρ(from graph) |
|---|---|---|---|
| $\rho_1$ | | | |
| $\rho_2$ | | | |
| $\rho_3$ | | | |

# EXPERIMENT 3 RESISTOR CIRCUITS AND WHEATSTONE BRIDGE

## OBJECTIVE

Resistor circuits with different combinations are investigated. Resultant of multi resistor circuit is calculated and the resultant value is measured by using simple DC circuits and applying Ohm's law. One of the very specific resistor combinations called Wheatstone-bridge circuit is investigated.

## THEORY AND PHYSICAL PRINCIPLES

Resistors can be connected in serial and in parallel in a circuit. Depending on the connection resultant or net effective resistance can be vastly different from the individual values of all connected resistances.

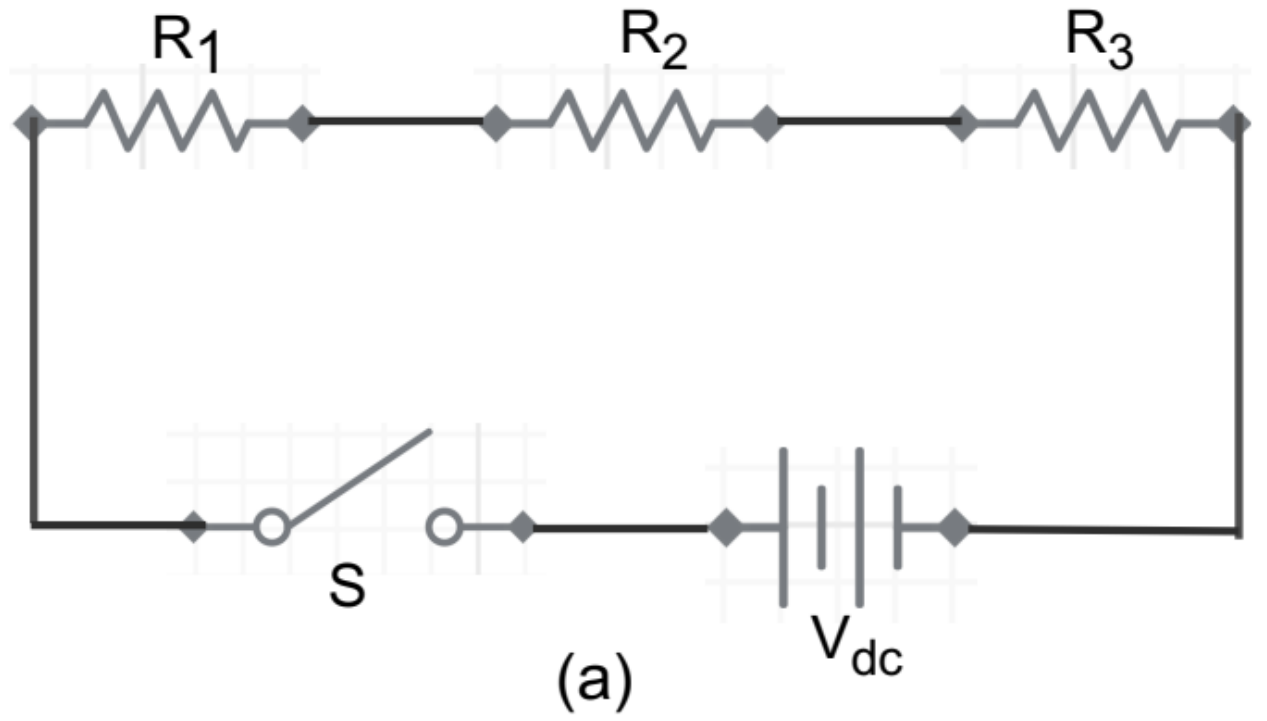

Figure 1 Series resistor circuit

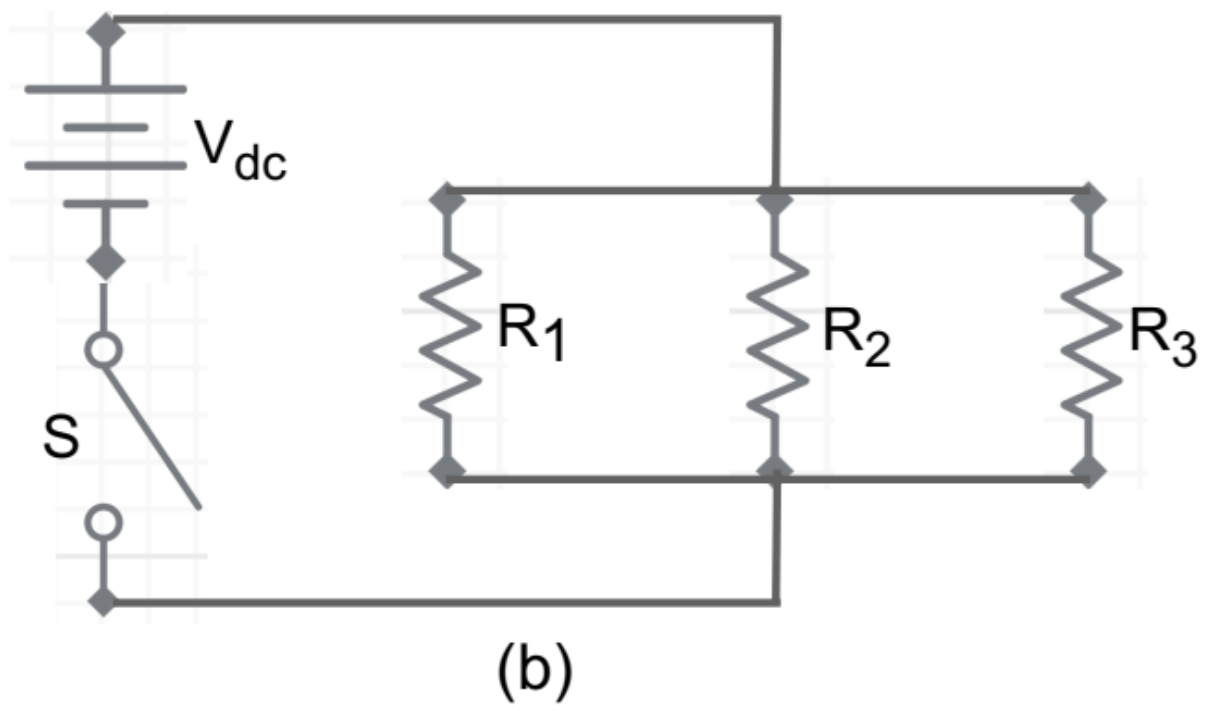

Figure 2 Parallel resistor circuit

Ohm's law for a resistor connected to DC voltage (battery),

$$V = IR \tag{1}$$

In serial circuits voltage segments along the line should be equal to the total voltage at the two end point of the line segment.

$$V = \sum V_i = V_1 + V_2 + V_3 \tag{2}$$

In parallel circuits current in every segment should be equal to the total current in the circuit.

$$I = \sum I_i = I_1 + I_2 + I_3 \tag{3}$$

When resistors are in series circuit, equivalent or resulting resistor is the addition of all the connected resistors and it can be found by using Ohm's law.

$$R_{equ} = \sum R_i = R_1 + R_2 + R_3 \tag{4}$$

When resistors are in parallel circuit, equivalent or resulting resistor is the inverse addition of all the connected resistors and it can be found by using Ohm's law.

$$R_{equ} = \left(\sum \frac{1}{R_i}\right)^{-1} = \left(\frac{1}{R_1} + \frac{1}{R_2} + \frac{1}{R_3}\right)^{-1} \tag{5}$$

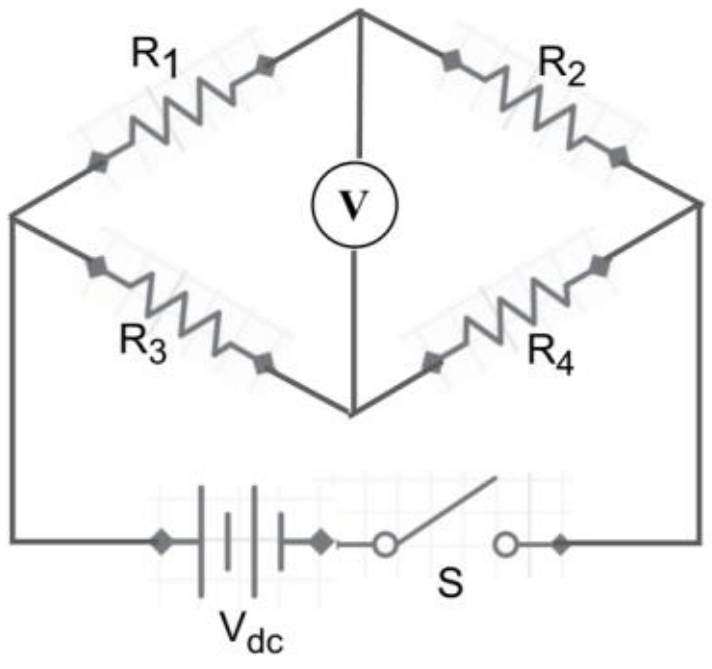

Figure 3 Wheatstone-bridge

Wheatstone-bridge is the resistor circuit with four different resistors as shown in figure 3. By using a variable resistor in the circuit, it is possible to get zero potential difference across mid points of the two parallel connections in the circuit.

When the Wheatstone bridge is balance,

$$\frac{R_1}{R_2} = \frac{R_3}{R_4} \tag{6}$$

## APPARATUS AND PROCEDURE

- Resistor circuit experiment is done with following simulation: https://phet.colorado.edu/en/simulation/circuit-construction-kit-dc
- A very detail video lesson of virtual lab (data collection with simulator and data analysis with excel) can be found here: https://youtu.be/vymVYbDfTbg

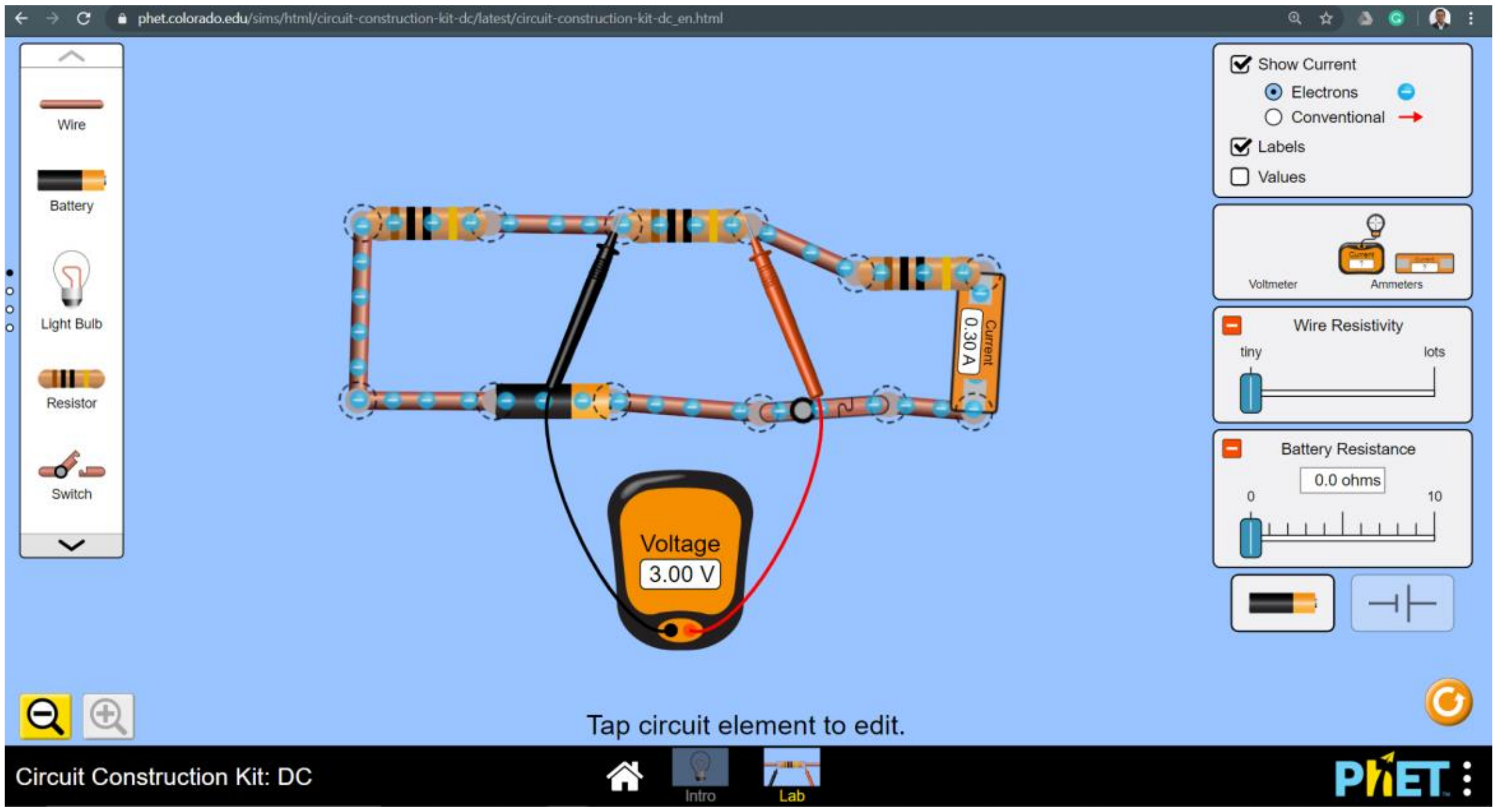

Figure 4 Simulation for resistor circuits (Picture credit: https://phet.colorado.edu)

- Build resistor circuits according to the given data tables.
- Calculate the equivalent resistor for each combination of serial, parallel and mixed combination of resistors.
- Measure current and voltage for each circuit.
- Complete tables of resistor circuit analysis.
- Build a Wheatstone-bridge circuit by using the given information in table-6.
- Balance the Wheatstone-bridge circuit by changing the resistor R4.
- Complete table 6 for two different types of Wheatstone-bridge circuits.
- Included the short procedure for each resistor circuit.

## PRE LAB QUESTIONS

1) When more resistors are added to the series circuit what does happen to the equivalent resistance?
2) When more resistors are added to the parallel circuit what does happen to the equivalent resistance?
3) Describe the current and voltage across each resistor in series circuit?
4) Describe the current and voltage across each resistor in parallel circuit?

## POST LAB QUESTIONS

1) What happens to the power through the circuit when more resistors are added to a series circuit? Explain your answer?
2) What happens to the power through the circuit when more resistors are added to a parallel circuit? Explain your answer?
3) When a Wheatstone-bridge is balanced what are the current and voltage through the resistor in the middle of the bridge?

## DATA ANALYSIS AND CALCULATIONS

*Resistors in Series and in Parallel*

Note the actual resistance.

$R_1$ = $R_2$ = $R_3$ =

a) Calculate total resistance for the following series combinations. Include a picture of each circuit diagram with all components.

Table 1 Series resistor combinations

| Series resistor Circuit | Current pass through the circuit [ ] | Voltage across each resistor [ ] | Resistor values measured (by V and I) [ ] | Equivalent resistance measured $R_{eq_meas}$ [ ] | Equivalent resistance Calculated $R_{eq_cal}$ [ ] | PD Between $R_{eq_cal}$ and $R_{eq_meas}$ |
|---|---|---|---|---|---|---|
| $R_1$ and $R_2$ | | | | | | |
| | | | | | | |
| $R_1$, $R_2$ and $R_3$ | | | | | | |
| | | | | | | |
| | | | | | | |

b) Calculate total resistance for the following parallel combinations. Include a picture of each circuit diagram with all components.

Table 2 Parallel resistor combinations

| Parallel resistor Circuit | Voltage through the circuit [ ] | Current across each resistor [ ] | Resistor values measured (by V and I) [ ] | Equivalent resistance measured $R_{eq_meas}$ [ ] | Equivalent resistance Calculated $R_{eq_cal}$ [ ] | PD Between $R_{eq_cal}$ and $R_{eq_meas}$ [ ] |
|---|---|---|---|---|---|---|
| $R_1$ and $R_2$ | | | | | | |
| | | | | | | |
| $R_1$, $R_2$ and $R_3$ | | | | | | |
| | | | | | | |
| | | | | | | |

c) Calculate total resistance for the following combined series/parallel combinations. Include a picture of each circuit diagram with all components.

Table 3 Analysis of mixed combinations of resistors

| Resistor Circuit | Voltage through the circuit [ ] | Current across each resistor [ ] | Resistor values measured (by V and I) [ ] | Equivalent resistance measured $R_{eq_meas}$ [ ] | Equivalent resistance Calculated $R_{eq_cal}$ [ ] | PD Between $R_{eq_cal}$ and $R_{eq_meas}$ [ ] |
|---|---|---|---|---|---|---|
| $R_1$, $R_2$ series and $R_3$ parallel | | | | | | |
| | | | | | | |
| | | | | | | |
| $R_1$, $R_3$ series and $R_2$ parallel | | | | | | |
| | | | | | | |
| | | | | | | |

d) Investigation of Wheatstone-bridge circuit (WBC). Include a picture of each circuit diagram with all components.

Table 4 Analysis of Wheatstone-bridge circuit

| Wheatstone-bridge circuit | Calculated $R_{4_cal}$ [ ] | Observed $R_{4_obs}$ [ ] | PD $R_{4_cal}$ and $R_{4_obs}$ |
|---|---|---|---|
| $R_1 = 25\ \Omega$,<br>$R_2 = 68\ \Omega$<br>$R_3 = 37\ \Omega$<br>$R_4 = X\ \Omega$<br>$R_4$ is unknown | | | |
| $R_1 = 25\ \Omega$<br>$R_2 = 56\ \Omega$<br>$R_3 = 45\ \Omega$<br>$R_4 = X\ \Omega$<br>$R_4$ is unknown | | | |
| $R_1 = 84\ \Omega$<br>$R_2 = 48\ \Omega$<br>$R_3 = 67\ \Omega$<br>$R_4 = X\ \Omega$<br>$R_4$ is unknown | | | |

# EXPERIMENT 4
# CAPACITOR PROPERTIES AND CONNECTIONS

## OBJECTIVE

Capacitor properties are investigated. Capacitance as a function of plate separation, plate area and dielectric constant are investigated. Capacitor connections are investigated. Equivalent or resultant of a multi capacitor circuit is calculated and checked with simulated results.

## THEORY AND PHYSICAL PRINCIPLES

Ohm's law for a resistor connected to DC voltage (battery),

$$V = IR \tag{1}$$

V is voltage, I is current, and R is resistance.

If a capacitor connected to a external voltage, then total charge stored in the capacitor,

$$Q = CV \tag{2}$$

Q is charge, C is capacitance and V is voltage of capacitor.

Capacitance of a capacitor depends on the geometric parameters of the capacitor.

$$C = \frac{\varepsilon_0 A}{d} \tag{3}$$

A is cross section area, d is plate separation of capacitor and $\varepsilon_0$ is permittivity of air (or free space).

When a dielectric medium is inserted into the capacitor, capacitance increases.

$$K = \frac{\varepsilon}{\varepsilon_0} = \frac{C}{C_0} \tag{4}$$

K is called dielectric constant and $\varepsilon$ is permittivity of medium.

Energy stored in capacitor,

$$U = \frac{1}{2}QV = \frac{1}{2}CV^2 = \frac{1}{2}\frac{Q^2}{C} \tag{5}$$

Energy density (energy per unit volume) of capacitor,

$$u = \frac{Energy}{volume} = \frac{1}{2}\varepsilon_0 E^2 \tag{6}$$

E is an electric field inside the capacitor plates.

If many capacitors are connected to a parallel circuit then each capacitor stores different amounts of charge. When capacitors are in a serial circuit, each capacitor stores the same amount of charge because the same current passes through each of them. When capacitors are in parallel circuit, each capacitor stores a different amount of charge because different current pass through each of them.

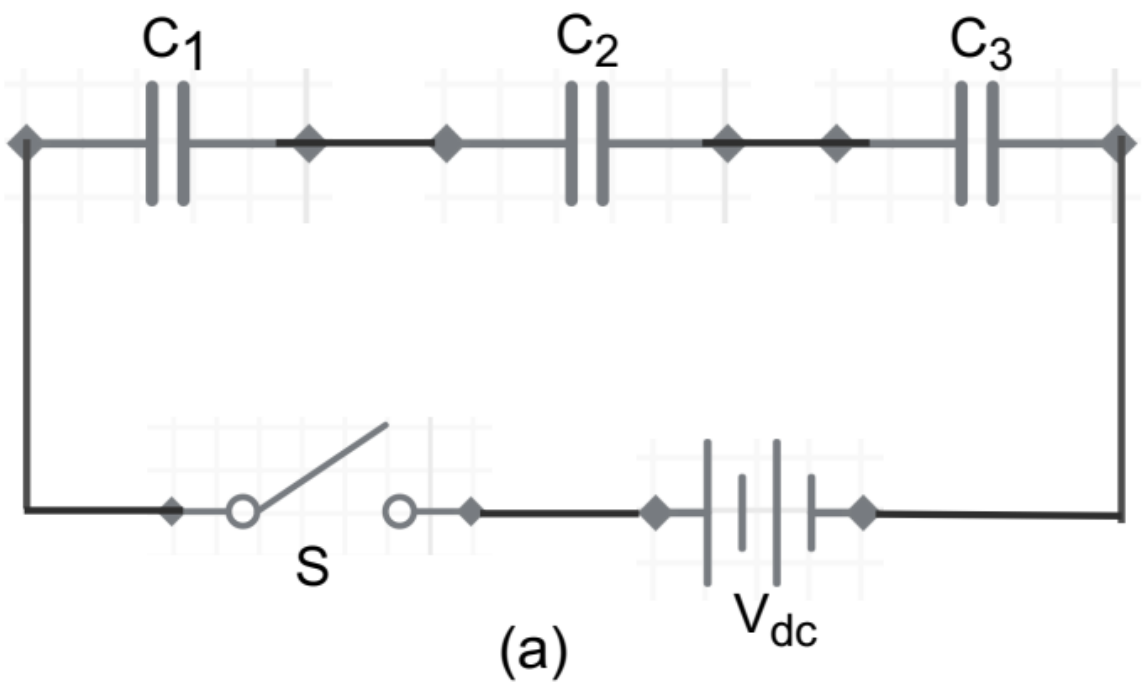

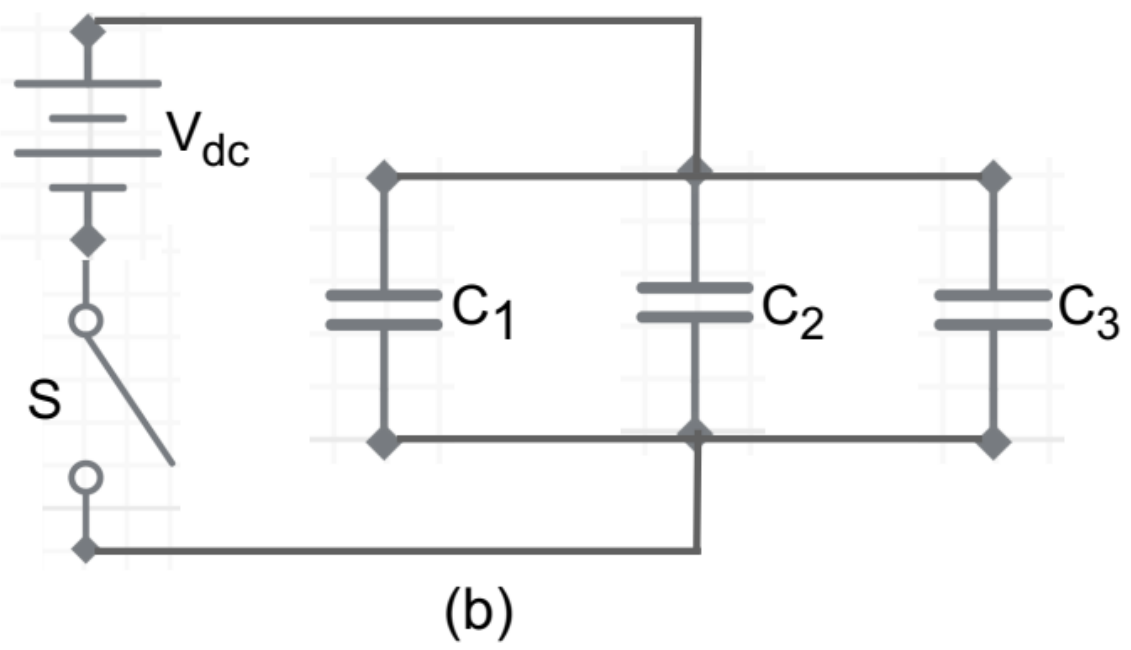

Figure 1 Series capacitor circuit

Figure 2 Parallel capacitor circuit

When capacitors are in series circuit, equivalent or resulting capacitor is the inverse addition of all the connected capacitors and it can be found by using capacitance equation.

$$C_{equ} = \left(\sum \frac{1}{C_i}\right)^{-1} = \left(\frac{1}{C_1} + \frac{1}{C_2} + \frac{1}{C_3}\right)^{-1} \qquad (8)$$

When capacitors are in parallel circuit, equivalent or resulting capacitor is the addition of all the connected capacitors and it can be found by using capacitance equations.

$$C_{equ} = \sum C_i = C_1 + C_2 + C_3 \qquad (9)$$

## APPARATUS AND PROCEDURE

- Capacitor circuit experiment is done with following simulation: https://phet.colorado.edu/sims/cheerpj/capacitor-lab/latest/capacitor-lab.html
- A very detail video lesson of virtual lab (data collection with simulator and data analysis with excel) can be found here: https://youtu.be/aE0YKsdOEBU

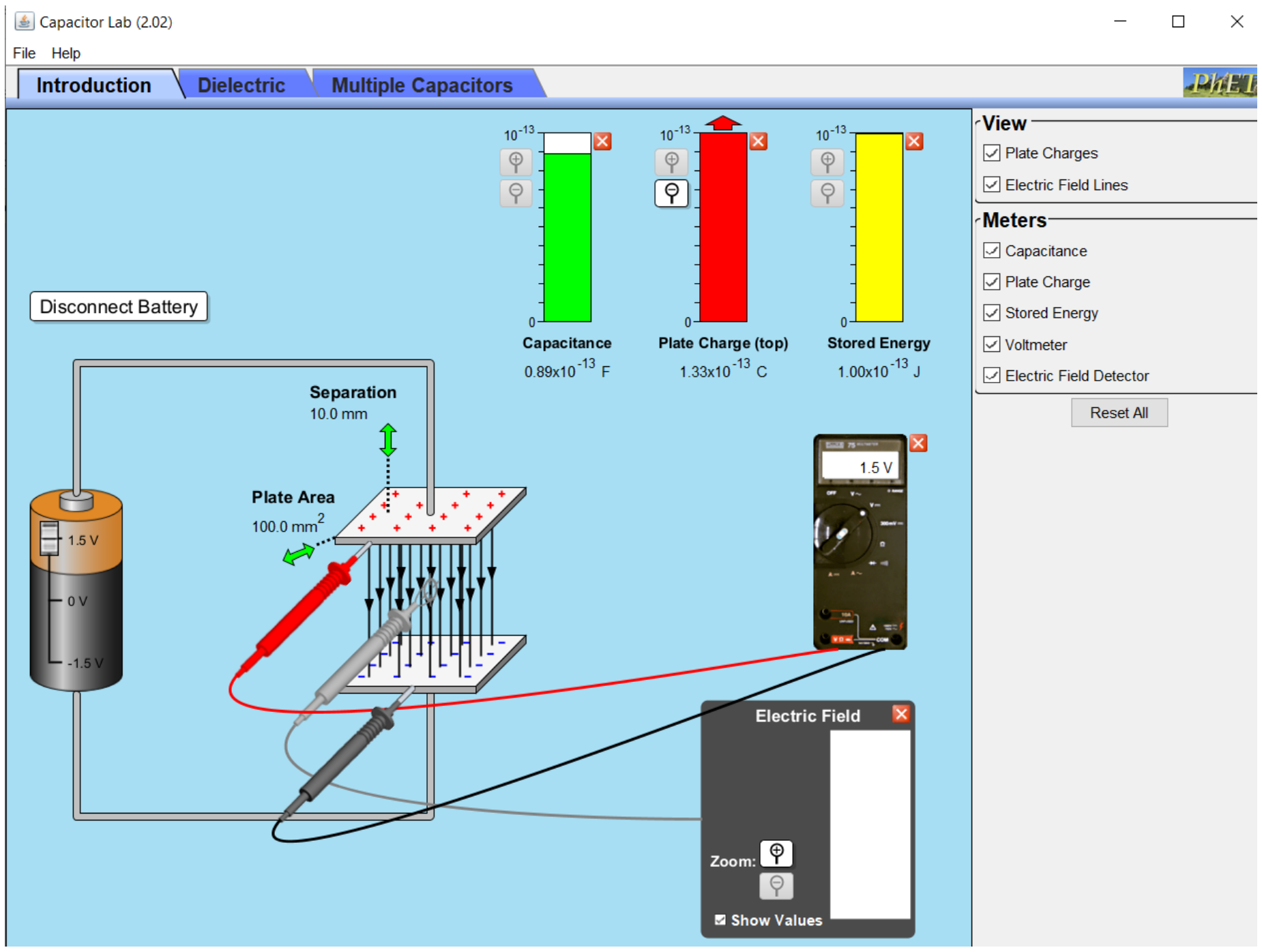

Figure 3 Simulation for capacitor circuit (Picture credit: https://phet.colorado.edu)

- Capacitance and electric field as a function of plate separation is studied in table-1 and 2.
- Fixed the plate area to the highest values in the simulator.
- Start the plate separation from the lowest value in the simulator and observe the capacitance and electric field between capacitor plates.
- Repeat the last two procedures by slowly increasing plate separation in steps till the table-1 and 2 is completed.
- Include the picture of this part in the procedure section.

- Capacitance as a function of plate area is studied in table-3.
- Fixed the plate separation to the lowest values in the simulator.
- Start the plate area from the lowest value in the simulator and observe the capacitance.
- Repeat the last two procedures by slowly increasing the plate area in steps till the table-3 is completed.
- Include the picture of this part in the procedure section.
- Effect dielectric medium inside the capacitor plates is studied in table-4.
- Fixed the plate separation into lowest value and the plate area into highest value in the simulator.
- Select the medium with lowest dielectric constant and insert it fully into the capacitor. Observe the capacitance and electric field inside the capacitance.
- Repeat the last procedure by changing the dielectric medium by using the selection tool in the right-hand side of the simulator.
- Include the picture of this part in the procedure section.

- Capacitor connections (series and parallel) are studied table-5.
- Calculate the equivalent capacitor for each combination of serial, parallel and mixed combination of capacitors.
- Observe and record the equivalent capacitance from the simulation.
- Include the picture of this part in the procedure section.

## PRE LAB QUESTIONS

1) Describe the capacitance of a capacitor?
2) Describe the electric field inside the capacitor?
3) Describe the effect of dielectric medium in capacitor?
4) Describe threshold breakdown voltage and electric field of capacitor?

## POST LAB QUESTIONS

1) When more capacitors are added into series circuits what happens to equivalent capacitance, and total energy stored?
2) When more capacitors are added into a parallel circuit what happens to equivalent capacitance, and total energy stored?

## DATA ANALYSIS AND CALCULATIONS

*Part A: Capacitor properties*

- Fixed the capacitor plate area into the highest value and then increased the plate serration from lowest to highest in steps.

- Area of the plate =

- Make a graph of C_cal vs plate separation. Explain the behavior by comparing it to the capacitance equation.

Table 1 Capacitance as a function of plate separation

| Plate separation d [ ] | Capacitance Observed $C_{obs}$ [ ] | Capacitance Calculated $C_{cal}$ [ ] | PD between $C_{obs}$ and $C_{cal}$ [ ] |
|---|---|---|---|
| | | | |
| | | | |
| | | | |
| | | | |
| | | | |

Table 2 Electric field analysis

| Plate separation d [ ] | Electric field Observed $E_{obs}$ [ ] | Energy density Observed $U_{obs}$ [ ] | Electric field Calculated $E_{cal}$ [ ] | PD between $E_{obs}$ and $E_{cal}$ [ ] |
|---|---|---|---|---|
| | | | | |
| | | | | |
| | | | | |
| | | | | |
| | | | | |

- Fixed the capacitor plate separation to lowest value and then increased the plate area from lowest to highest in steps.

- Plate separation =

- Make a graph of C_cal vs plate area. Explain the behavior by comparing it to the capacitance equations.

Table 3 Capacitance as a function of plate area

| Plate Area A [ ] | Capacitance Observed $C_{obs}$ [ ] | Capacitance Calculated $C_{cal}$ [ ] | PD between $C_{obs}$ and $C_{cal}$ [ ] |
|---|---|---|---|
| | | | |
| | | | |
| | | | |
| | | | |
| | | | |

*Part B: Capacitor properties with dielectric medium*

- Filed the capacitor plate area to A=400.0mm$^2$ and plate separation to d=10.0mm.
- Insert the dielectric medium into the capacitor slowly (about 1.0mm at a time).
- Note down the capacitance, charge stored, and energy stored.

Table 4 Capacitance and electric field with dielectric medium

| Dielectric medium with dielectric constant | Capacitance Observed $C_{obs}$ [ ] | Capacitance Calculated $C_{cal}$ [ ] | Electric field observed $E_{obs}$ [ ] | Electric field calculated $E_{cal}$ [ ] | PD between $C_{obs}$ and $C_{cal}$ [ ] | PD between $E_{obs}$ and $E_{cal}$ [ ] |
|---|---|---|---|---|---|---|
| | | | | | | |
| | | | | | | |
| | | | | | | |
| | | | | | | |

*Part C: Capacitor circuits (series and parallel)*

- Note the actual capacitance of the capacitors.

$C_1$ = $C_2$ = $C_3$ =

- Measure the total capacitance for the following series and parallel combinations. Include the picture of circuit diagram in procedure.

Table 5 Equivalent capacitor of combined capacitor circuits

| Capacitor circuit | Capacitance Observed $C_{obs}$ [ ] | Capacitance Calculated $C_{cal}$ [ ] | PD between $C_{obs}$ and $C_{cal}$ [ ] |
|---|---|---|---|
| Series $C_1$, $C_2$ and $C_3$ | | | |
| Parallel $C_1$, $C_2$, and $C_3$ | | | |
| $C_1$, $C_2$ series and parallel with $C_3$ | | | |
| $C_2$, $C_3$ parallel and series with $C_1$ | | | |

# EXPERIMENT 5 CHARGING AND DISCHARGING CAPACITOR

## OBJECTIVE

Capacitor charging and discharging circuit (RC circuit) is investigated by manually measuring the charging/discharging time and voltage change of the capacitor.

## THEORY AND PHYSICAL PRINCIPLES

Charging/discharging capacitor is studied by using a simple RC circuit. By measuring voltage and time during charging and discharging capacitor behavior in DC circuits can be studied.

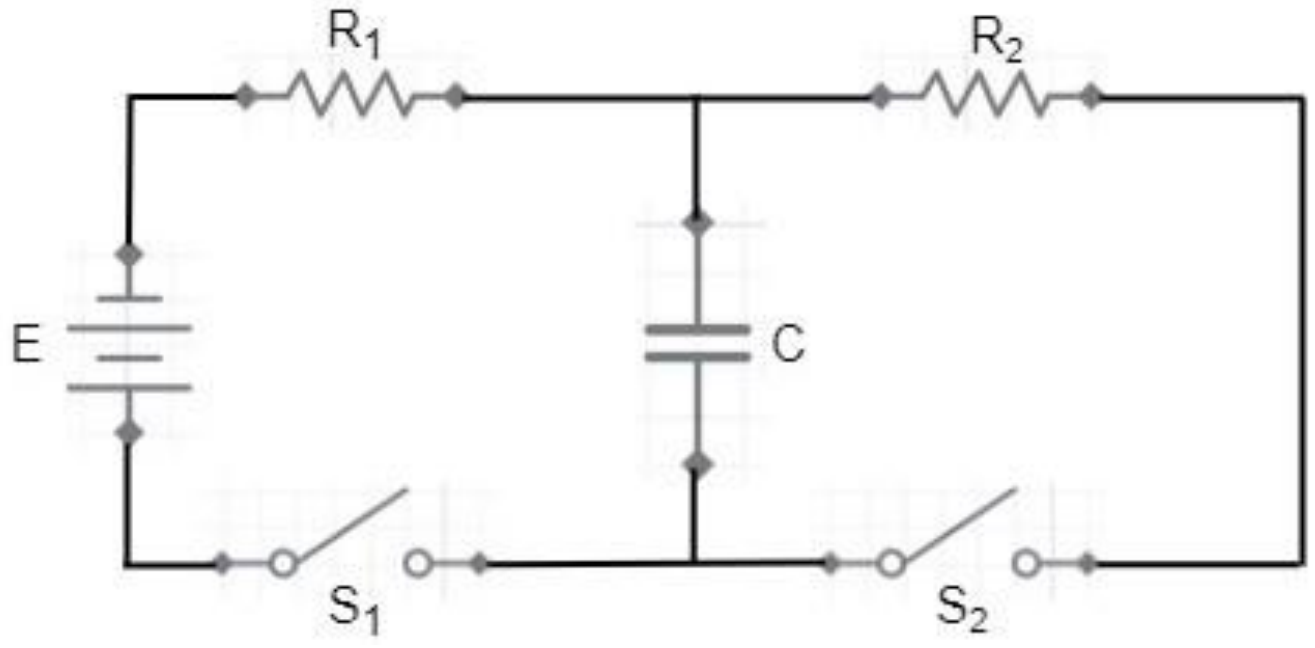

Figure 1 Charging and discharging capacitor circuit

Consider charging a capacitor circuit (left side loop in the figure 1) and apply Kirchhoff loop rule.

$$\sum E + \sum IR = 0 \tag{1}$$

$$E - IR - V_c = 0 \tag{2}$$

E is the battery voltage and R is the resistor.

Change the current ($I$) and capacitor voltage ($V_c$) in terms of charges in the capacitor at any given time ($t$),

$$E - R\frac{dq}{dt} - \frac{q}{C} = 0 \tag{3}$$

$$\frac{dq}{(EC-q)} = \frac{1}{RC}dt \tag{4}$$

Integrate equation (4) to find the charge build in capacitor in time ($t$),

$$\int_0^q \frac{dq}{(EC-q)} = \frac{1}{RC}\int_0^t dt \tag{5}$$

By solving equation (5),

$$q(t) = Q_0\left(1 - e^{-t/\tau}\right) \tag{6}$$

$Q_0$ is maximum possible charge stored and $\tau = RC$ is the time constant.

Voltage across charging capacitor in RC circuit,

$$V_c(t) = V_0\left(1 - e^{-t/\tau}\right) \tag{7}$$

$$\frac{V_C(t)}{V_0} = 1 - e^{-t/\tau} \tag{8}$$

$$ln\left(1 - \frac{V_c}{V_0}\right) = -\frac{1}{\tau}t \tag{9}$$

$\tau$ is called the time constant and $\tau = RC$ and $V_0$ maximum voltage.

R is resistance and C capacitance in a simple DC circuit. And Vo is maximum voltage which is supply voltage.

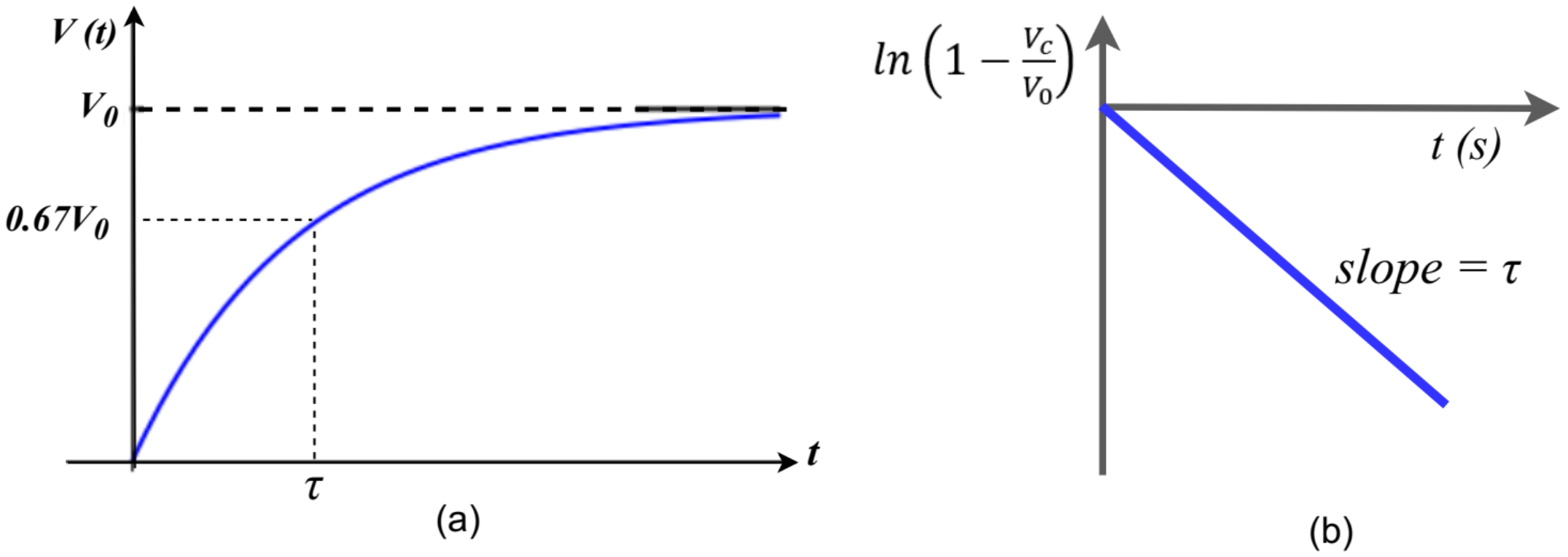

Figure 2 Charging capacitor, (a) Voltage as a function of time and (b) linearize behavior of voltage

Consider discharging the capacitor circuit (right side loop in the figure 1) and apply Kirchhoff loop rule.

$$\sum E + \sum IR = 0 \tag{10}$$

$$IR + V_c = 0 \tag{11}$$

Change the current ($I$) and capacitor voltage ($V_c$) in terms of charges in the capacitor at any given time ($t$),

$$R\frac{dq}{dt} + \frac{q}{C} = 0 \tag{12}$$

$$\frac{dq}{q} = \frac{1}{RC}dt \tag{13}$$

Integrate equation (13) to find the charge build in capacitor in time ($t$),

$$\int_q^o \frac{dq}{q} = \frac{1}{RC}\int_0^t dt \tag{14}$$

By solving equation (14), $q(t) = Q_0 e^{-t/\tau}$ (15)

Voltage across discharging capacitor,

$$V_c(t) = V_0 e^{-t/\tau} \tag{16}$$

$$ln\left(\frac{V_c}{V_0}\right) = -\frac{1}{\tau}t \tag{17}$$

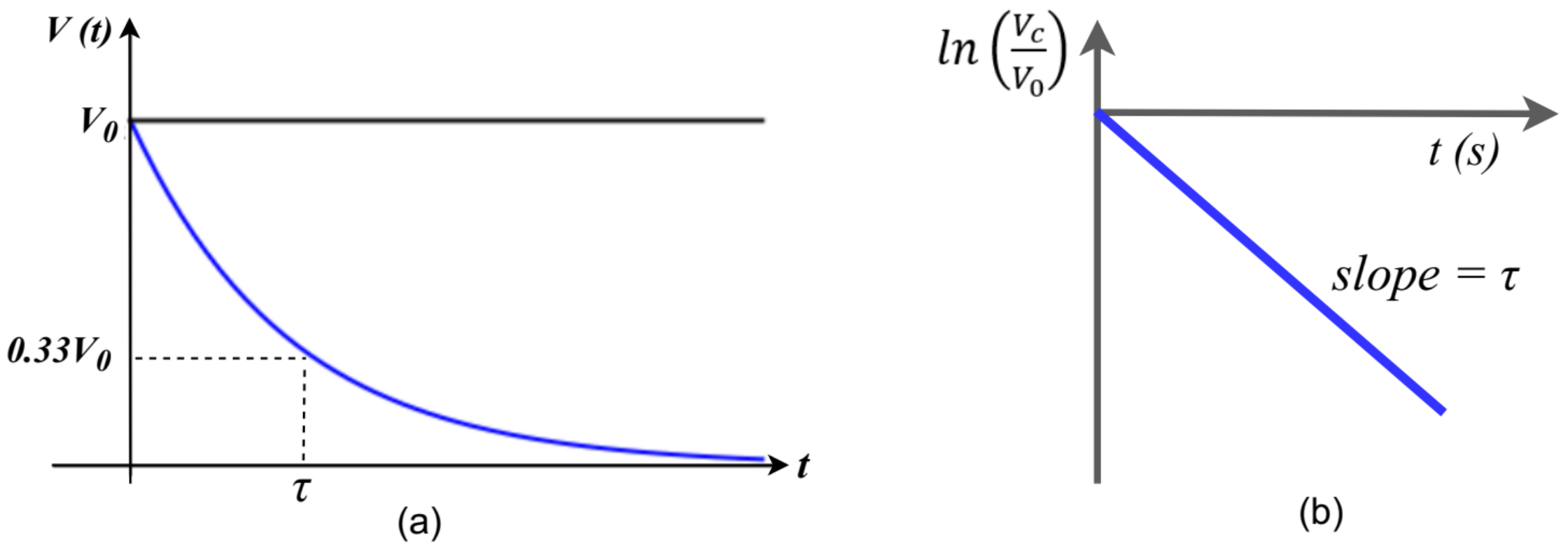

Figure 3 Discharging capacitor, (a) Voltage as a function of time and (b) linearize behavior of voltage.

## APPARATUS AND PROCEDURE

- Capacitor charging and discharging process is studied by using following simulation: http://phet.colorado.edu/sims/html/circuit-construction-kit-ac/latest/circuit-construction-kit-ac_en.html
- A very detail video lesson of virtual lab (data collection with simulator and data analysis with excel) can be found here: https://youtu.be/1_sudrTLU0U

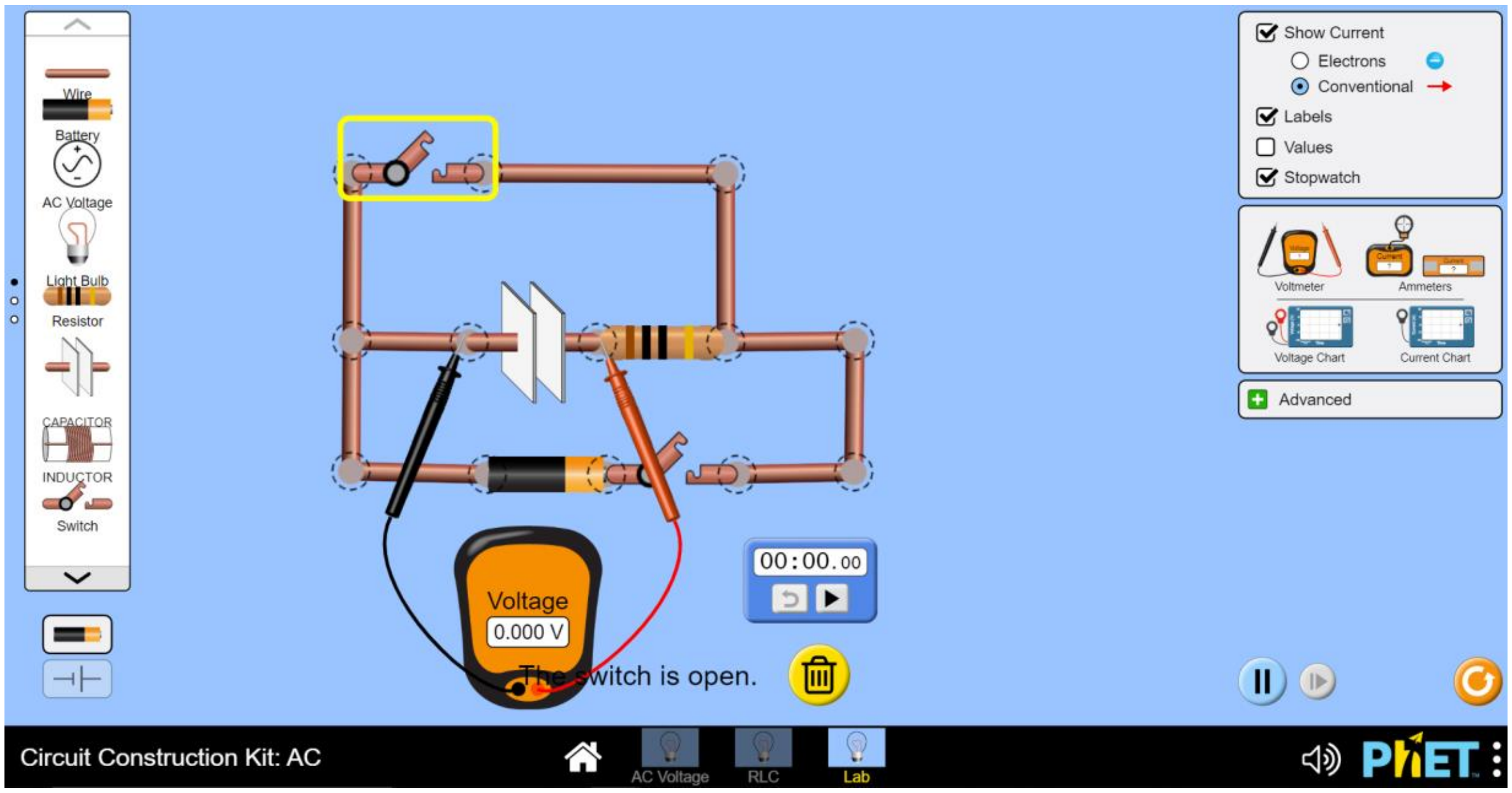

Figure 4 Simulation for RC circuit (Picture credit: https://phet.colorado.edu)

- Set up the circuit as shown in the figure 4 with following information, C=0.1000 F and R=20.0Ω.
- Bottom loop of the circuit simulates the charging capacitor and the top loop of the circuit simulates the discharging capacitor.
- Make sure the capacitor is fully discharged and then keep both switches open.
- Add a stopwatch to the simulation page by clicking the icon “stopwatch”.
- Make sure the play button of the simulator (button on the very bottom of the simulation page) is on pause mode.
- Then close the switch in the bottom loop (charging circuit is on) and make sure the switch in the top loop is open (the discharging circuit is off).
- Extract the voltage at each half a second by clicking the fast-forward button on the simulator. Fast-forward button changes the time in 0.1 seconds, and it is easy to get the exact voltage across the capacitor every 0.5 seconds.
- After the capacitor is fully charged, open the switch on the bottom loop (charging circuit) and close the switch on the top loop (discharging circuit), which simulates the discharging capacitor.
- Extract the voltage and at each half a second by using the fast forward button.

## PRE LAB QUESTIONS

1) Describe the time constant of RC circuit?
2) Describe the behavior of charging capacitors in DC circuits?
3) Describe the behavior of discharging capacitors in DC circuits?

## POST LAB QUESTIONS

1) Find the current and energy stored after one time constant in the charging capacitor?
2) Find the current and energy stored after one time constant in dis-charging capacitor?

## DATA ANALYSIS AND CALCULATIONS

*Charging and discharging capacitor*

A. *RC circuit with one resistor and one capacitor*

- Voltage and time of charging and discharging capacitor should be extracted from the simulator and include a picture of the circuit.

Table 1 Analysis of charging and discharging capacitor

| Charging Capacitor | | | Discharging Capacitor | | |
|---|---|---|---|---|---|
| Time [ ] | Voltage [ ] | $ln\left(1-\frac{V_c}{V_0}\right)$ | Time [ ] | Voltage [ ] | $ln\left(\frac{V_c}{V_0}\right)$ |
| | | | | | |
| | | | | | |
| | | | | | |
| | | | | | |
| | | | | | |
| | | | | | |
| | | | | | |
| | | | | | |
| | | | | | |
| | | | | | |
| | | | | | |
| | | | | | |
| | | | | | |
| | | | | | |
| | | | | | |
| | | | | | |

- Make a graph of voltage vs time for the charging capacitor to observe the behavior.
- Then make a graph of $ln\left(1-\frac{V_c}{V_0}\right)$ vs time and fit the data with linear fitting.
- Find the time constant ($\tau_1$) by using the slope of the graph for charging capacitor and compare it with expected time constant ($\tau$=RC).
- Make a graph of voltage vs time for the discharging capacitor to observe the behavior.
- Then make a graph of $ln\left(\frac{V_c}{V_0}\right)$ vs time and fit the data with linear fitting.
- Find the time constant ($\tau_2$) by using the slope of the graph discharging capacitor and compare it with expected time constant ($\tau$=RC).

B. *RC circuit with one resistor and two serially connected capacitors*

- Voltage and time of charging and discharging capacitor should be extracted from the simulator and include a picture of the circuit.

Table 2 Analysis of charging and discharging capacitor

| Charging Capacitor | | | Discharging Capacitor | | |
|---|---|---|---|---|---|
| Time [ ] | Voltage [ ] | $ln\left(1-\frac{V_c}{V_0}\right)$ | Time [ ] | Voltage [ ] | $ln\left(\frac{V_c}{V_0}\right)$ |
| | | | | | |
| | | | | | |
| | | | | | |
| | | | | | |
| | | | | | |
| | | | | | |
| | | | | | |
| | | | | | |
| | | | | | |
| | | | | | |
| | | | | | |
| | | | | | |
| | | | | | |
| | | | | | |
| | | | | | |
| | | | | | |

- Make a graph of voltage vs time for the charging capacitor to observe the behavior.
- Then make a graph of $ln\left(1-\frac{V_c}{V_0}\right)$ vs time and fit the data with linear fitting.
- Find the time constant ($\tau_1$) by using the slope of the graph for charging capacitor and compare it with expected time constant ($\tau$=RC).
- Make a graph of voltage vs time for the discharging capacitor to observe the behavior.
- Then make a graph of $ln\left(\frac{V_c}{V_0}\right)$ vs time and fit the data with linear fitting.
- Find the time constant ($\tau_2$) by using the slope of the graph discharging capacitor and compare it with expected time constant ($\tau$=RC).

*C. RC circuit with one resistor and two parallel connected capacitors*

- Voltage and time of charging and discharging capacitor should be extracted from the simulator and include a picture of the circuit.

Table 3 Analysis of charging and discharging capacitor

| Charging Capacitor | | | Discharging Capacitor | | |
|---|---|---|---|---|---|
| Time [ ] | Voltage [ ] | $ln\left(1-\frac{V_c}{V_0}\right)$ | Time [ ] | Voltage [ ] | $ln\left(\frac{V_c}{V_0}\right)$ |
| | | | | | |
| | | | | | |
| | | | | | |
| | | | | | |
| | | | | | |
| | | | | | |
| | | | | | |
| | | | | | |
| | | | | | |
| | | | | | |
| | | | | | |
| | | | | | |
| | | | | | |
| | | | | | |
| | | | | | |
| | | | | | |

- Make a graph of voltage vs time for the charging capacitor to observe the behavior.
- Then make a graph of $ln\left(1-\frac{V_c}{V_0}\right)$ vs time and fit the data with linear fitting.
- Find the time constant ($\tau_1$) by using the slope of the graph for charging capacitor and compare it with expected time constant ($\tau$=RC).
- Make a graph of voltage vs time for the discharging capacitor to observe the behavior.
- Then make a graph of $ln\left(\frac{V_c}{V_0}\right)$ vs time and fit the data with linear fitting.
- Find the time constant ($\tau_2$) by using the slope of the graph discharging capacitor and compare it with expected time constant ($\tau$=RC).

# EXPERIMENT 6 MULTILOOP CIRCUIT AND KIRCHHOFF'S RULES

## OBJECTIVE

Multiloop DC circuit is investigated. Kirchhoff's loop rule and junction rule are investigated by measuring voltage and current in the multiloop circuit.

## THEORY AND PHYSICAL PRINCIPLE

When a circuit contains more than one loop then voltage and current across each segment of the multiloop can be investigated by using Kirchhoff's rules.

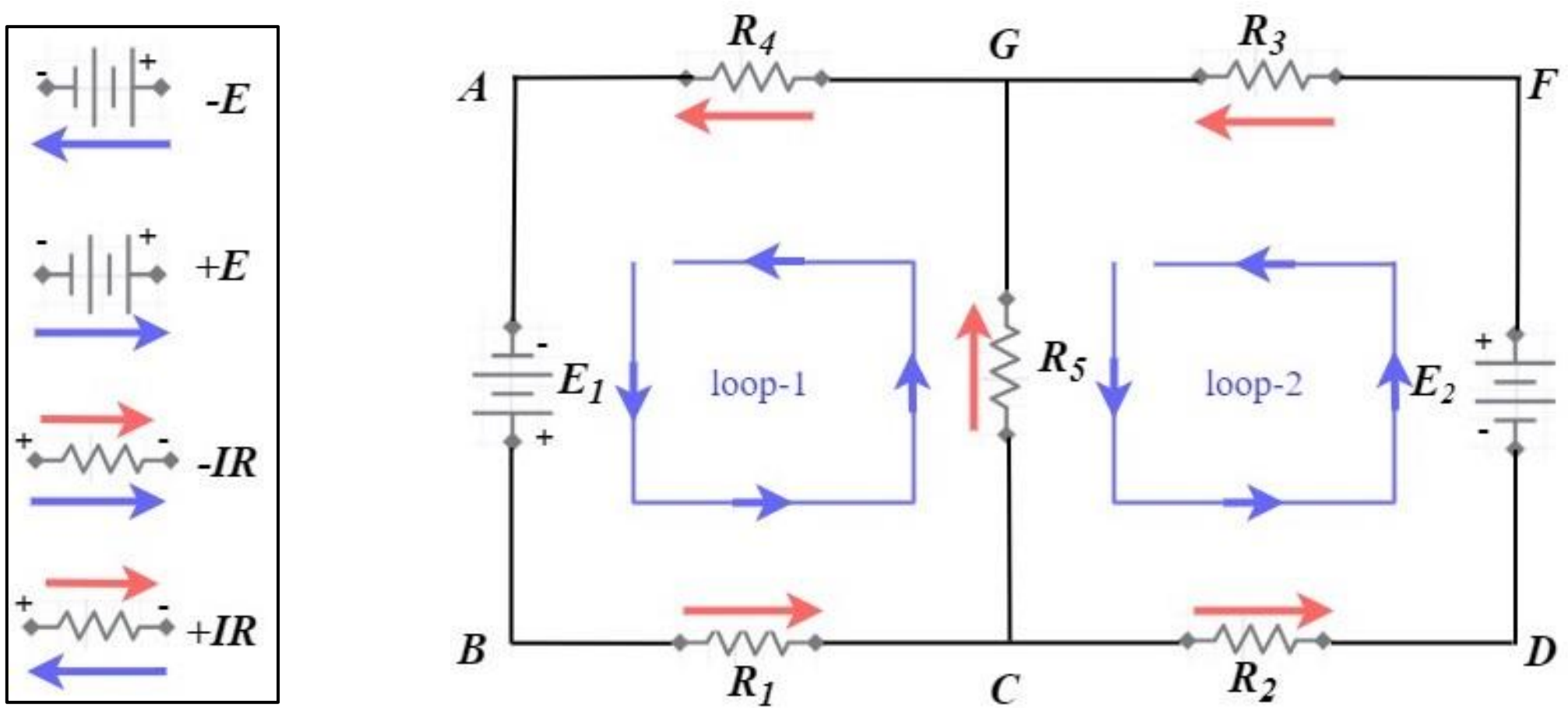

$V_1 = 9.00\ V$

$E_1$ =$V_1$= Battery

$E_2 = 6.00\ V$

$E_2$ =$V_2$= Battery

$R_1 = 18.0\ \Omega$

$R_2 = 14.0\ \Omega$

$R_3 = 22.0\ \Omega$

$R_4 = 28.0\ \Omega$

**(a)** **(b)** **(c)**

Figure 1 (a) Sign conventions, (b) multiloop circuit and (c) battery voltages and resistor values. Red arrows represent current direction and blue arrows represent the loop-direction.

Kirchhoff's rules consist of two important rules namely the junction rule and loop-rule. Junction rule states that the current coming into a junction must leave the junction.

$$I_{IN} = I_{OUT} \tag{1}$$

Applying to junction C,

$$I_1 = I_2 + I_3 \tag{2}$$

When applying the junction rule to C and G, both produce the same equation.

Kirchhoff's loop-rule states that the voltage added into a closed (battery or DC power supply) must be equal to the sum of all the voltage drop through the loop.

$$\sum E + \sum IR = 0 \tag{3}$$

Loop-rule must apply with sign rules which can be found on figure-1(a). To apply loop-rule a closed loop must be selected in multi-loop circuit.

Consider loop-1 (ABCGA),

$$V_1 - I_1R_1 - I_3R_5 - I_1R_4 = 0 \tag{4}$$

Consider loop-2 (DFGCD),

$$V_2 - I_2R_3 + I_3R_5 - I_2R_2 = 0 \tag{5}$$

Consider outer loop (ABCDFGA),

$$V_1 + V_1 - I_1R_1 - I_2R_2 - I_2R_3 - I_1R_4 = 0 \tag{6}$$

By solving questions 2, 4, 5, it is possible to find the current pass through each segment of the above multiloop circuit.

## APPARATUS AND PROCEDURE

- This experiment is done with the following simulation: https://phet.colorado.edu/sims/html/circuit-construction-kit-dc-virtual-lab/latest/circuit-construction-kit-dc-virtual-lab_en.html
- A very detail video lesson of virtual lab (data collection with simulator and data analysis with excel) can be found here: https://youtu.be/YDdQloZmIqM

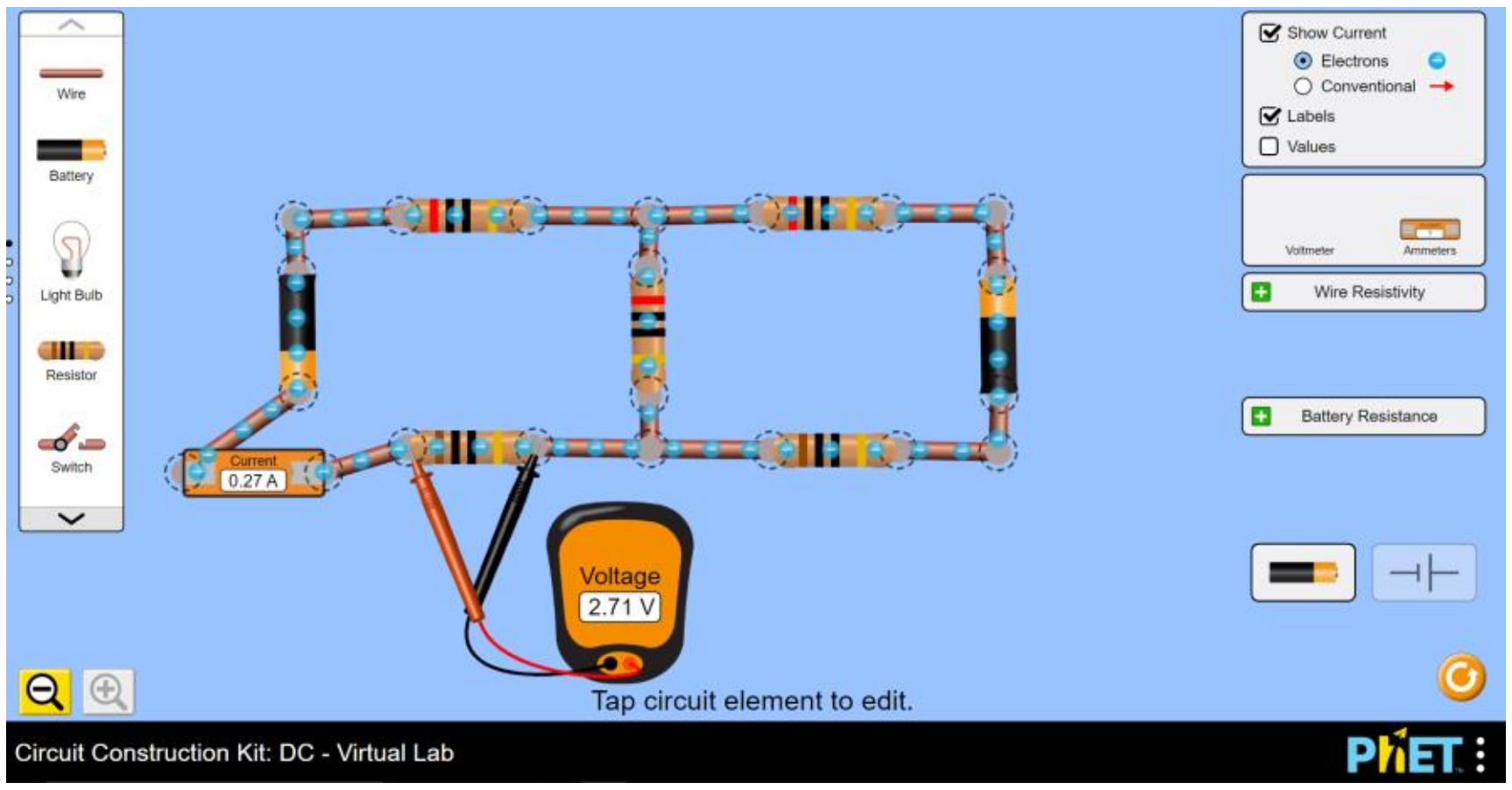

Figure 2 Multiloop circuit with the simulation (Photo credit: https://phet.colorado.edu)

- First click the above link and then make the multiloop circuit in the figure 1.
- Measure current through each resistor by using an ammeter which must be connected serial to the loop.
- Measure voltage through each resistor by using voltmeter which must be connected parallel with resistor. Also, the positive probe of the voltmeter must be connected relative to the positive end of the battery.

## PRE LAB QUESTIONS

1) Describe Kirchhoff's rules?
2) Describe sign rules for resistor and battery?

## POST LAB QUESTIONS

1) Calculate total power supplied and total power dissipation on loop-1 in the circuit in figure-1(b)?
2) Calculate total power supplied and total power dissipation on loop-2 in the circuit in figure-1(b)?
3) Calculate total power supplied and total power dissipation on the whole circuit in figure-1?
4) Does the power supply and power dissipation in each loop and the whole circuit are the same in the above calculation? Explain your answer?

## DATA ANALYSIS AND CALCULATIONS

A. *Current analysis in the multiloop circuit by Kirchhoff's rules and direct measurements*

- *Write down three equations (1 from junction-rule and 2 from loop-rule) by using Kirchhoff's rules. Then find currents $I_1$ (GABC), $I_2$ (CDFG) and $I_3$ (CG) by solving above linear equations.*

Table 1 Current analysis in the multiloop circuit

| Linear equations from Kirchhoff's rules | Current in the multiloop circuit | | | |
|---|---|---|---|---|
| | Symbol | Calculated values [ ] | Measured values [ ] | Percent difference |
| | $I_1$ | | | |
| | $I_2$ | | | |
| | $I_3$ | | | |

B. *Verifying Kirchhoff's junction rule*

Table 2 Verifying junction rule

| *Method* | *Total current into the junction* $\sum I_{in}$ [ ] | *Total current out from the junction* $\sum I_{out}$ [ ] | *Percent difference* |
|---|---|---|---|
| *Calculated* | | | |
| *Measured* | | | |

C. *Voltage analysis by calculation and direct measurements. PD should be done without considering the sign on measured voltage.*

Table 3 Analysis of voltage on loop-1 (ABCGA)

| Voltage | Voltage Calculated $V_{cal}=IR$ [ ] | Voltage Measured [ ] | Percent difference [ ] |
|---|---|---|---|
| $V(R_1)$ | | | |
| $V(R_4)$ | | | |
| $V(R_5)$ | | | |

Table 4 Analysis of voltage on loop-2 (DFGCD)

| Voltage | Voltage Calculated $V_{cal}=IR$ [ ] | Voltage Measured [ ] | Percent difference [ ] |
|---|---|---|---|
| $V(R_2)$ | | | |
| $V(R_3)$ | | | |
| $V(R_5)$ | | | |

Table 5 Analysis of voltage on outer loop (ABCDFGA)

| Voltage | Voltage Calculated $V_{cal}=IR$ [ ] | Voltage Measured [ ] | Percent difference [ ] |
|---|---|---|---|
| $V(R_1)$ | | | |
| $V(R_2)$ | | | |
| $V(R_3)$ | | | |
| $V(R_4)$ | | | |

*D. Verifying Kirchhoff's loop rule*

Table 6 Verifying loop rule

| *Method* | *Total voltage supply into the loop* $\sum E$ [ ] | *Total voltage drop across the loop calculated* $\sum IR$ [ ] | *Total voltage drop across the loop measured* $\sum V$ [ ] | *Percent difference between voltage supply and voltage drop(calculated)* | *Percent difference between voltage supply and voltage drop(measured)* |
|---|---|---|---|---|---|
| *Loop-1 ABCGA* | | | | | |
| *Loop-2 DFGCD* | | | | | |
| *Outer Loop ABCDFGA* | | | | | |

# EXPERIMENT 7 SOURCES OF MAGNETIC FIELD

## OBJECTIVE

Magnetic field due to current carrying wire is investigated. Permeability constant is calculated by using the equation of magnetic field of an infinitely long straight wire. Magnetic field of a solenoid is investigated.

## THEORY AND PHYSICAL PRINCIPLES

Current carrying wire produces magnetic field loops around the wire and it is observed that the magnetic field strength is a function of current and distance from the wire. Magnetic field produced around current carrying wire can be found by using Biot-Savart law, which is the most fundamental law that shows how to find the magnetic field around infinitely small pieces of current wire.

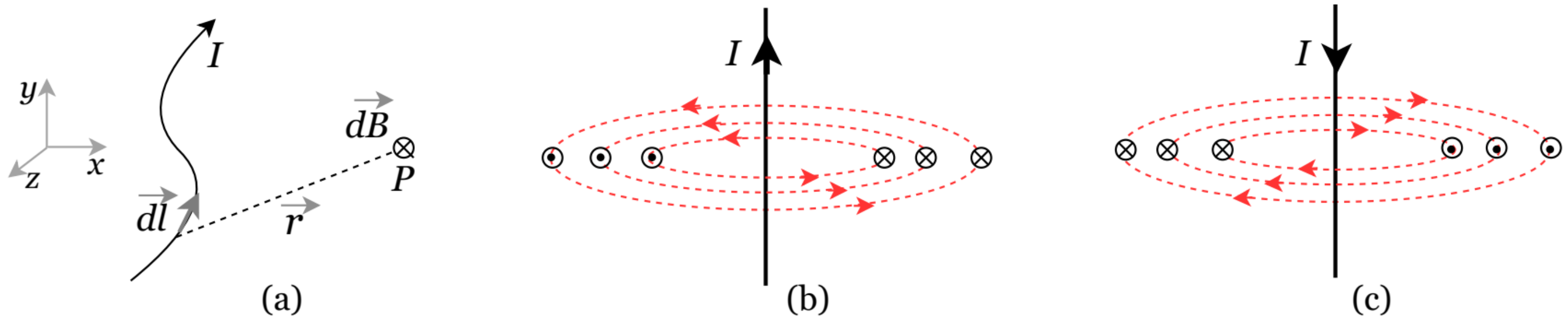

Figure 1 (a) Small length of wire $\vec{dl}$ with current $I$ produce and small magnetic field $\vec{dB}$ at $\vec{r}$ distance, (b) infinitely long straight wire with current in +y direction and magnetic field around the wire in xz plane, (c) infinitely long straight wire with current in -y direction and magnetic field around the wire in xz plane

Biot-Savart law, $\vec{dB} = \frac{\mu_0 I}{4\pi}\frac{\vec{dl}\otimes\hat{r}}{r^2}$ (1)

$\vec{dB}$ is magnetic field at point P, $\vec{dl}$ is small piece of wire length, $r$ is displacement from wire piece to point P, $I$ is current in the wire, $\hat{r}$ is the unit vector of the displacement vector $r$.

$\mu_0 = 4\pi \times 10^{-7}\frac{Tm}{A}$, which is called the permeability of free space (or air).

Magnetic fields due to infinitely long straight current wire can be found by using Biot-Savart law.

$$B = \frac{\mu_0 I}{2\pi r} \quad (2)$$

$$B \propto I \text{ and } B \propto \frac{1}{r} \quad (3)$$

Magnetic field is directly proportional to current, and inversely proportional with distance.

Direction of the magnetic field around the wire is shown in figure 1(b and c). When the current is in +y direction and the magnetic field produces in xz plane and the field loops are in counter-clock-wise direction. When the current direction changes to -y then the magnetic field loop direction changes into clock-wise direction. Magnetic field direction around current wire can be found by using right-hand-rule (RHR); when the thumb of the right hand is pointed into the direction of current then the rotation of the other four fingers of the right hand shows the direction of the magnetic field around the wire.

When the current is rotated into a cylindrical shape as shown in figure 2(a) it is called solenoid. Since the solenoid consists of many numbers of current loops, the magnetic field of a solenoid can be found by using Ampere's law.

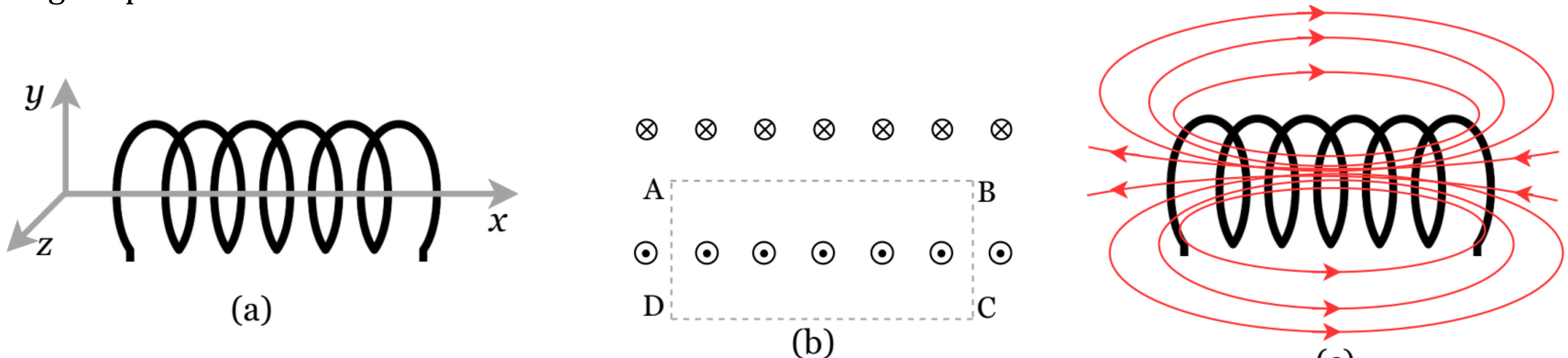

Figure 2 (a) solenoid with length is on x axis and each wire loop is on yz plane, (b) cross section view of solenoid on xy plane, (c) magnetic field lines of solenoid

Ampere's law shows an easy way to find the magnetic field due to collection of current wires. When define the close loop around a current wire (or collection of current wires),

$$\oint \vec{B} \cdot \vec{dl} = \mu_0 I_{enc} \tag{4}$$

B is the magnetic field on a small length *dl* in close Ampere loop, $I_{enc}$ is the sum of all current inside the close Ampere loop.

Figure 2(b) shows the close Ampere loop (ABCDA) for solenoid.

$$\oint \vec{B} \cdot \vec{dl} = \int_A^B \vec{B}_{AB} \cdot \vec{dl} + \int_B^C \vec{B}_{BC} \cdot \vec{dl} + \int_C^D \vec{B}_{CD} \cdot \vec{dl} + \int_D^A \vec{B}_{DA} \cdot \vec{dl} \tag{5}$$

Due to symmetry and the current direction, $\vec{B}_{BC}$, $\vec{B}_{CD}$ and $\vec{B}_{DA}$ becomes zero or the magnitude of extremely small. N is the number of turns along the line segment of AB and I is in each loop.

$$\oint \vec{B} \cdot \vec{dl} = \int_A^B \vec{B}_{AB} \cdot \vec{dl} = \mu_0 I_{enc} = \mu_0 NI \tag{6}$$

If assume the solenoid is very long (infinitely long approximation), $\vec{B}_{AB}$ is constant along the line segment of AB. n is called the loop density of the solenoid.

$$\int_A^B \vec{B}_{AB} \cdot \vec{dl} = B \int_A^B dl = BL = \mu_0 NI \tag{7}$$

$$B = \mu_0 \frac{N}{L} I = \mu_0 nI \tag{8}$$

## APPARATUS AND PROCEDURE

*A. Magnetic field due to infinitely long straight wire*

- This experiment is performed by using computer simulation. Please click here to access the simulation: http://cdac.olabs.edu.in/?sub=74&brch=9&sim=90&cnt=4
- A very detail video lesson of virtual lab (data collection with simulator and data analysis with excel) can be found here: https://youtube.com/playlist?list=PLsVLYnCPRO5mTootqIOgLUXzfohxdao5c

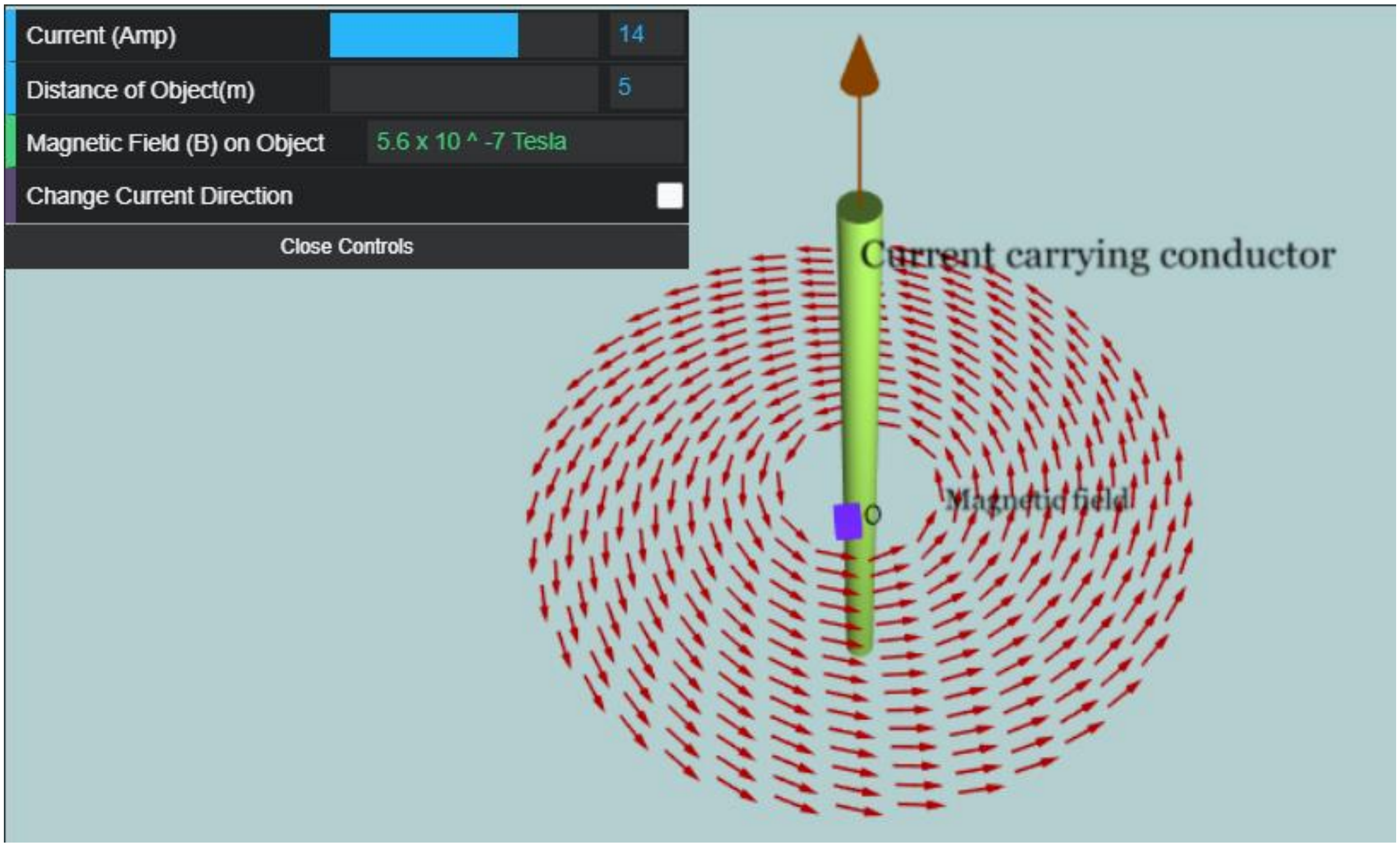

Figure 3 Simulation for magnetic field around current wire (Picture credit: http://cdac.olabs.edu.in)

- Set the current to 5.0A.
- Set the object location to 1.0m position. This will measure the magnetic field 1.0m away from the wire.
- Note the radial distance, magnetic field and direction of magnetic field lines.
- Then flip the current direction in the wire and observe the magnetic field and direction of field lines.

- Slowly increase the object location and note down the radial distance and magnetic field at every 5.0m increment.
- For each case estimate the permeability constant by using measure radial distance and magnetic field.
- Find the average of calculated permeability constant.
- Find the percent error between average calculated and known values of permeability constants.
- Make a graph of magnetic field vs radial distance (r) and discuss the behavior of the graph.
- Make a graph of magnetic field vs 1/r and discuss the behavior of the graph.

- Then set the radial distance at 5.0m.
- Note down the magnetic field with increasing current into the wire.
- Make a graph of magnetic field vs current and discuss the behavior of the graph.

*B. Magnetic field inside Solenoid*

- This experiment is performed by using computer simulation. Please click here to access the simulation: http://cdac.olabs.edu.in/?sub=74&brch=9&sim=91&cnt=4
- A very detail video lesson of virtual lab (data collection with simulator and data analysis with excel) can be found here: https://youtube.com/playlist?list=PLsVLYnCPRO5mTootqIOgLUXzfohxdao5c

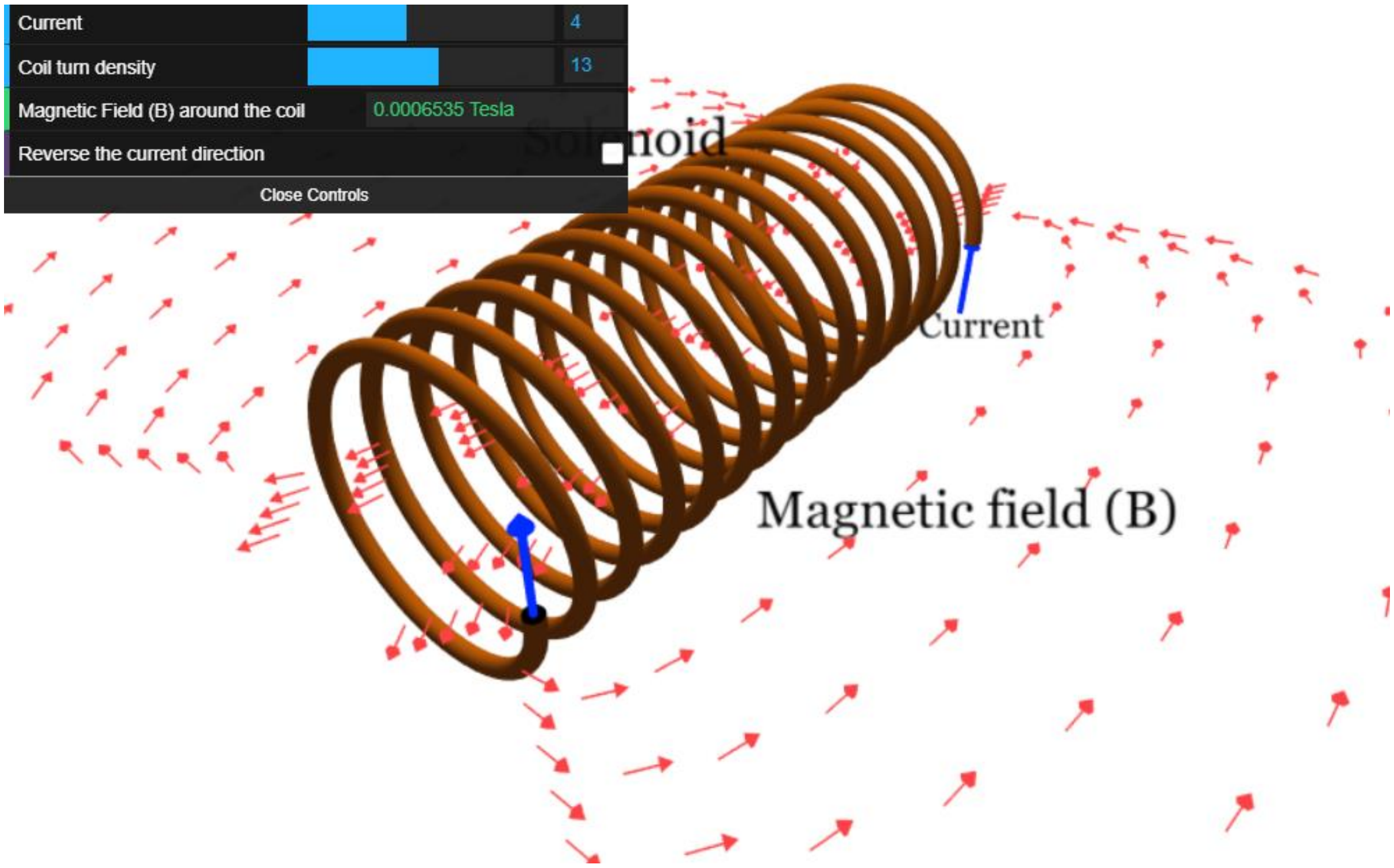

Figure 4 Simulation for magnetic induction inside solenoid (Picture credit: http://cdac.olabs.edu.in)

- Set the current and coil turn disunity into highest values.
- Observes the current direction relative to the center axis of solenoid.
- Identify magnetic poles at the ends of the solenoid.

- Set the coil turn density to number 6.
- Set the current to 2.0A.
- Note down the magnetic field inside the solenoid with increasing current of 1.0A at a time. Keep the coil turn density fixed.
- Fixed the current at 10.0A.
- Note down the magnetic field inside the solenoid with increasing coil turn density by 2 at a time. Keep the current fixed at 10.0A.
- Make a graph of magnetic field vs current and discuss the behavior of the graph.
- Make a graph of magnetic field vs coil turn density and discuss the behavior of the graph.

## PRE LAB QUESTIONS

1) Describe magnetic field behavior of a current carrying wire?
2) Describe right-hand-rule for current wire?
3) Describe Ampere's law?
4) Describe the magnetic field of a solenoid?

**DATA ANALYSIS AND CALCULATIONS**

*A. Magnetic field due to infinitely long straight wire*

Table 1 Magnetic field and its direction due to current wire

| Current direction | Current I [ ] | Radial distance R [ ] | Magnetic field B [ ] | Magnetic field direction [ ] |
|---|---|---|---|---|
| | | | | |
| | | | | |

Table 2 Magnetic field of current wire at constant current

| Current I [ ] | Magnetic field B [ ] | Permeability constant $\mu_0$(cal) [ ] | PE between $\mu_0$(cal) and $\mu_0$ [ ] |
|---|---|---|---|
| | | | |
| | | | |
| | | | |
| | | | |
| | | | |
| | | | |
| | | | |
| | | | |
| | | | |
| | | | |

- *Find the percent error between average calculated and known values of permeability constants.*
- *Make a graph of magnetic field vs radial distance (r) and discuss the behavior of the graph.*
- *Make a graph of magnetic field vs 1/r and discuss the behavior of the graph.*

Table 3 Magnetic field of current wire at constant radial distance

| Radial distance<br>R<br>[ ] | Magnetic field<br>B<br>[ ] | Permeability constant $\mu_0$(cal)<br>[ ] | PE between $\mu_0$(cal) and $\mu_0$<br>[ ] |
|---|---|---|---|
| | | | |
| | | | |
| | | | |
| | | | |
| | | | |
| | | | |
| | | | |
| | | | |
| | | | |
| | | | |

- *Make a graph of magnetic field vs current and discuss the behavior of the graph.*

*C. Magnetic field inside Solenoid*

Table 4 Magnetic pole of a solenoid

| Current direction | Current<br>I<br>[ ] | Coil turn density<br>n<br>[ ] | Magnetic field<br>[ ] | Magnetic pole in left of solenoid | Magnetic pole in right of solenoid |
|---|---|---|---|---|---|
| | | | | | |
| | | | | | |

Table 5 Magnetic field of solenoid at constant coil density

| Coil turn density = | | Current = | |
|---|---|---|---|
| Current<br>I<br>[ ] | Magnetic field<br>B<br>[ ] | Coil turn density<br>n<br>[ ] | Magnetic field<br>B<br>[ ] |
| | | | |
| | | | |
| | | | |
| | | | |
| | | | |
| | | | |
| | | | |
| | | | |
| | | | |
| | | | |

- Make a graph of magnetic field vs current and discuss the behavior of the graph.
- Make a graph of magnetic field vs coil turn density and discuss the behavior of the graph.

# EXPERIMENT 8 ELECTROMAGNETIC INDUCTION

## OBJECTIVES

The effect of rate of change of magnetic field is investigated by using various methods.

- Magnetic field due to bar magnetic and the magnetic field strength as a function of strength of bar magnet.
- Faraday and Lenz laws: Rate of change of magnetic field will induce electromotive force.
- Electromagnet and induction.
- Applications of EM-induction: Transformer and generator.

## THEORY AND PHYSICAL PRINCIPLES

More than two centuries ago (1819), Danish scientist Hans Christian Oersted discovered one of the most important experimental evidence in history which is the first evidence of the magnetic field produced due to current. When a compass is placed near by the current carrying wire, the compass needle deflects due to current. This observation was led to further investigation to test the possibility of the electric current due to the magnetic field. It was first observed by Joseph Henry in England and Michael Faraday in America in 1831. They conducted an experiment by using a galvanometer and permanent magnet which provide the first evidence of electromagnetic induction and it is generally known as Faraday law of induction. This observation led to some of the particularly important technological developments such as voltage transformers, electric power generation, and other application electronic circuits.

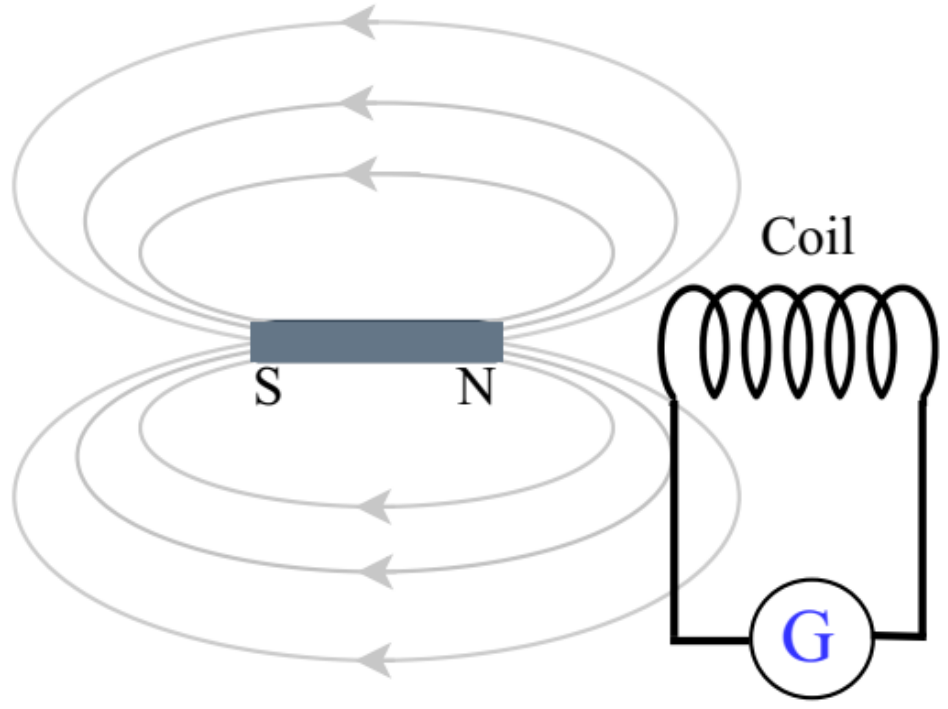

Figure 1 Electromagnetic induction

When permanent magnet moves towards the solenoid it is observed that the galvanometer deflects. Direction of galvanometer deflection explains in Lenz's law which named after Henrich Lenz which states that the induced current direction must oppose the change which produced it. This further confirms by the galvanometer deflection changes when the magnet moves into the coil and it moves out of the coil. Lenz law explains that the electromagnetic circuits agrees with Newton laws and conservation of energy.

Electromagnetic induction explains by combining Faraday and Lenz's laws as follows, induced emf (E) is equal to the negative of rate of change of magnetic flux.

Magnetic flux, $\emptyset = \int \vec{B} \cdot d\vec{A} = BAcos\theta$ (1)

Induced emf, $\varepsilon = -N\frac{d\emptyset}{dt} = -\frac{d(BAcos\theta)}{dt}$ (2)

$$\varepsilon = -N\left(Acos\theta\frac{dB}{dt} + Bcos\theta\frac{dA}{dt} + BA\frac{dcos\theta}{dt}\right) \quad (3)$$

Magnetic field (B), cross section area (A), angle between magnetic field and the surface unit vector ($\theta$), number of loops in the coil (N).

## APPARATUS AND PROCEDURE

- This experiment is performed completely by using computer simulation. Please click here to access the simulation: https://phet.colorado.edu/sims/cheerpj/faraday/latest/faraday.html?simulation=faraday
- A very detail video lesson of virtual lab (data collection with simulator and data analysis with excel) can be found here: https://youtu.be/hTinRCrCejE

*A. Magnetic field lines around bar magnet*

- Magnetic field lines around the bar magnet are investigated by using the first option in the simulation.
- Set the simulation as given in the figure below.
- Check the magnetic field behavior around the bar magnet by changing the strength.

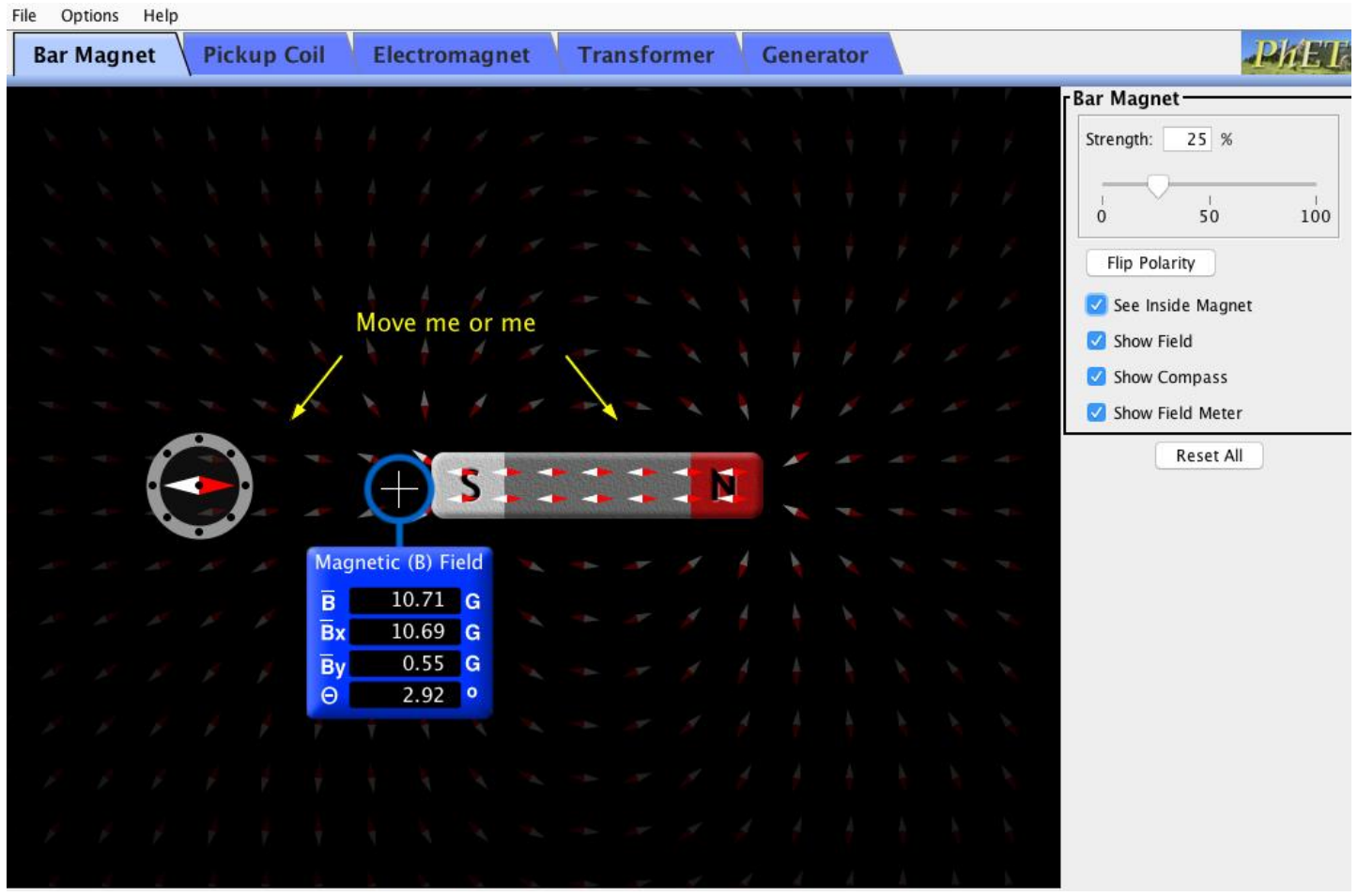

Figure 2 Simulation for magnetic field due to bar-magnet (Picture credit: https://phet.colorado.edu/)

*B. Using stronger bar-magnet to induce current in a Solenoid*

- Use the second option (pickup coil) in the simulation as shown in the figure.
- Move the north pole of the magnet in and out of the coil. Record the observations in table-1.
- Move the south pole of the magnet in and out of the coil. Record the observations in table-2.
- Explain the observations in both tables in your report in terms of Faraday and Lenz laws?

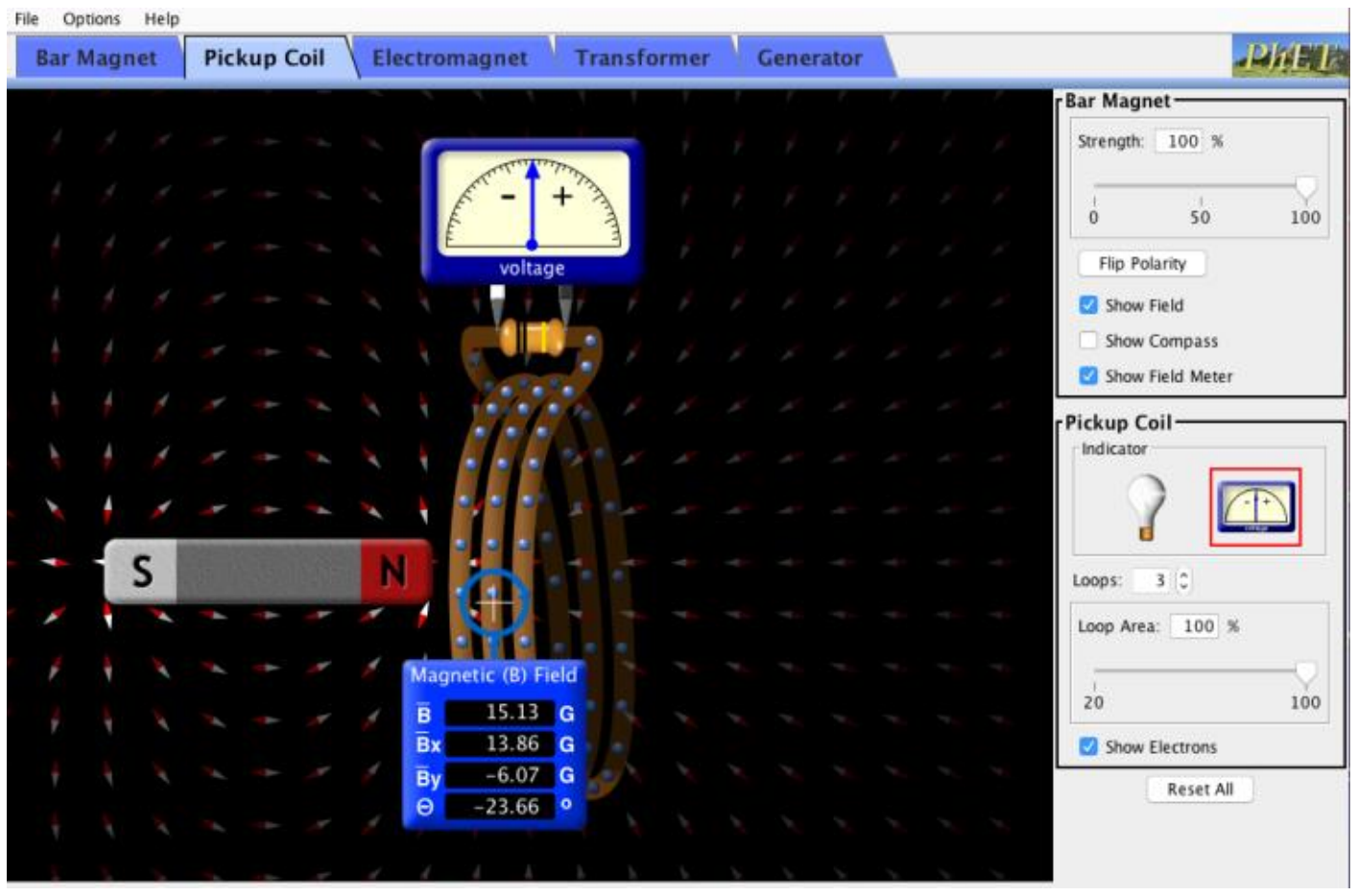

Figure 3 Simulation for magnetic induction due to bar-magnet (Picture credit: https://phet.colorado.edu)

*C. Induced magnetic field due to current in the solenoid*

- Use the third option (electromagnet) in the simulation as shown in the figure.
- Place the magnetic field sensor and compass very close to the coil as shown in the figure.
- Set the battery voltage to zero.
- Increase battery voltage slowly 1.0 V at a time and record the observed direction of magnetic field and the measured strength of the field in table-3.
- Discuss the results in table-3 in your report.

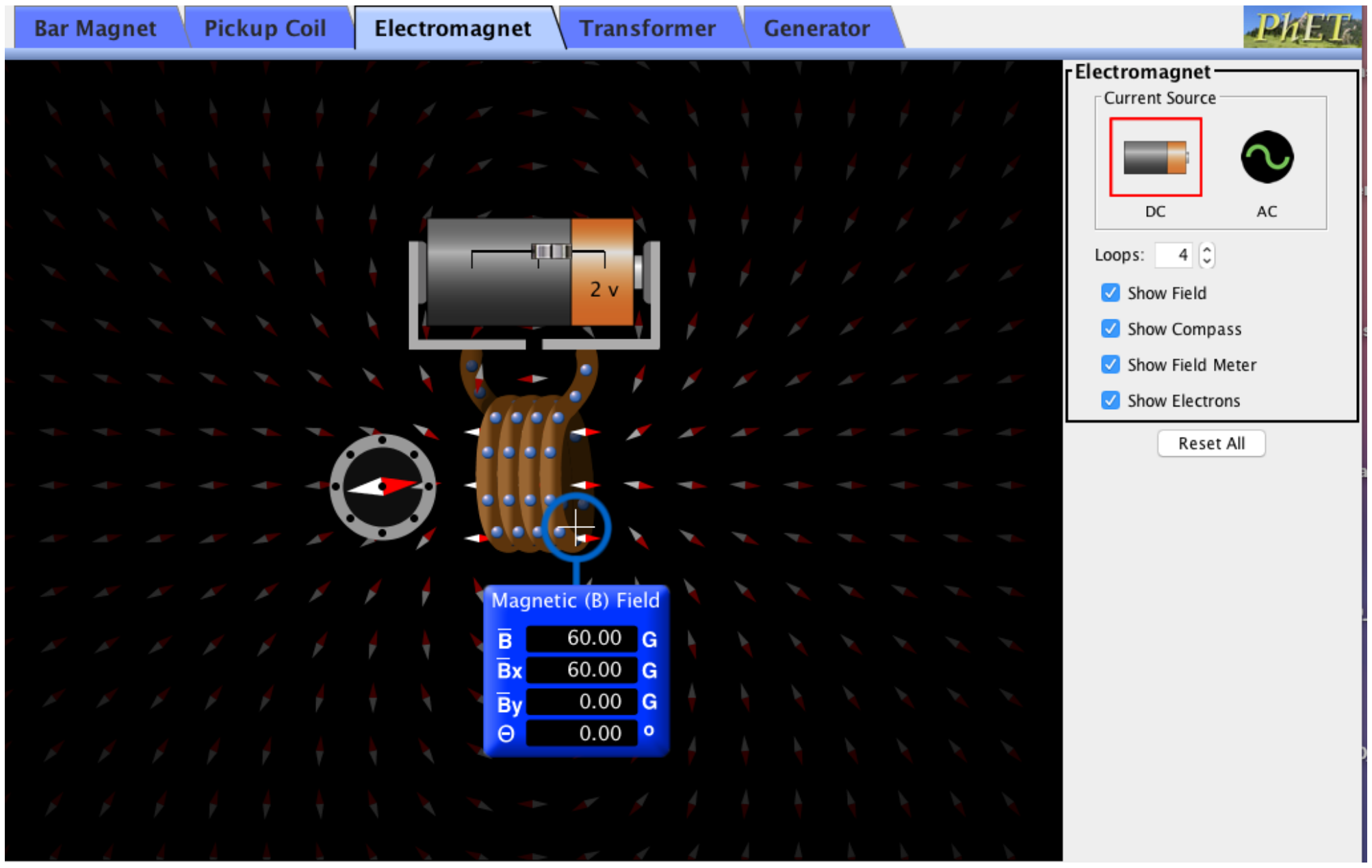

Figure 4 Simulation for electromagnet (Picture credit: https://phet.colorado.edu/)

*D. Changing current in one solenoid to induce current in the secondary solenoid*

- Use the fourth option (transformer) in the simulation as shown in the figure.
- Set up the two solenoids as shown in the figure. One with a galvanometer (pickup coil) and the other with a battery (electromagnet).
- In this experiment the aim is to measure the induced emf in the pickup coil. Emf in secondary coil can be only observed when the magnetic flux changes inside the coil-2. This can be done by switching on and off the battery.
- Set the loop area of the secondary coil to 20%.
- Click and drag the battery voltage switch quickly from zero to maximum voltage in a positive direction.
- Observe the deflection (direction and value) of the galvanometer in the secondary coil.
- Record the observation on table-4.
- Then repeat above last 3 of the above procedure for negative direction of the battery voltage.
- Repeat the last five processes of the above procedure by increasing the cross section area of the secondary coil by 20% at a time.
- Discuss the results in terms of Faraday and Lenz laws in table-4 on your report.
- Also discuss the behavior of induced emf in the secondary coil as s function of its cross section area.

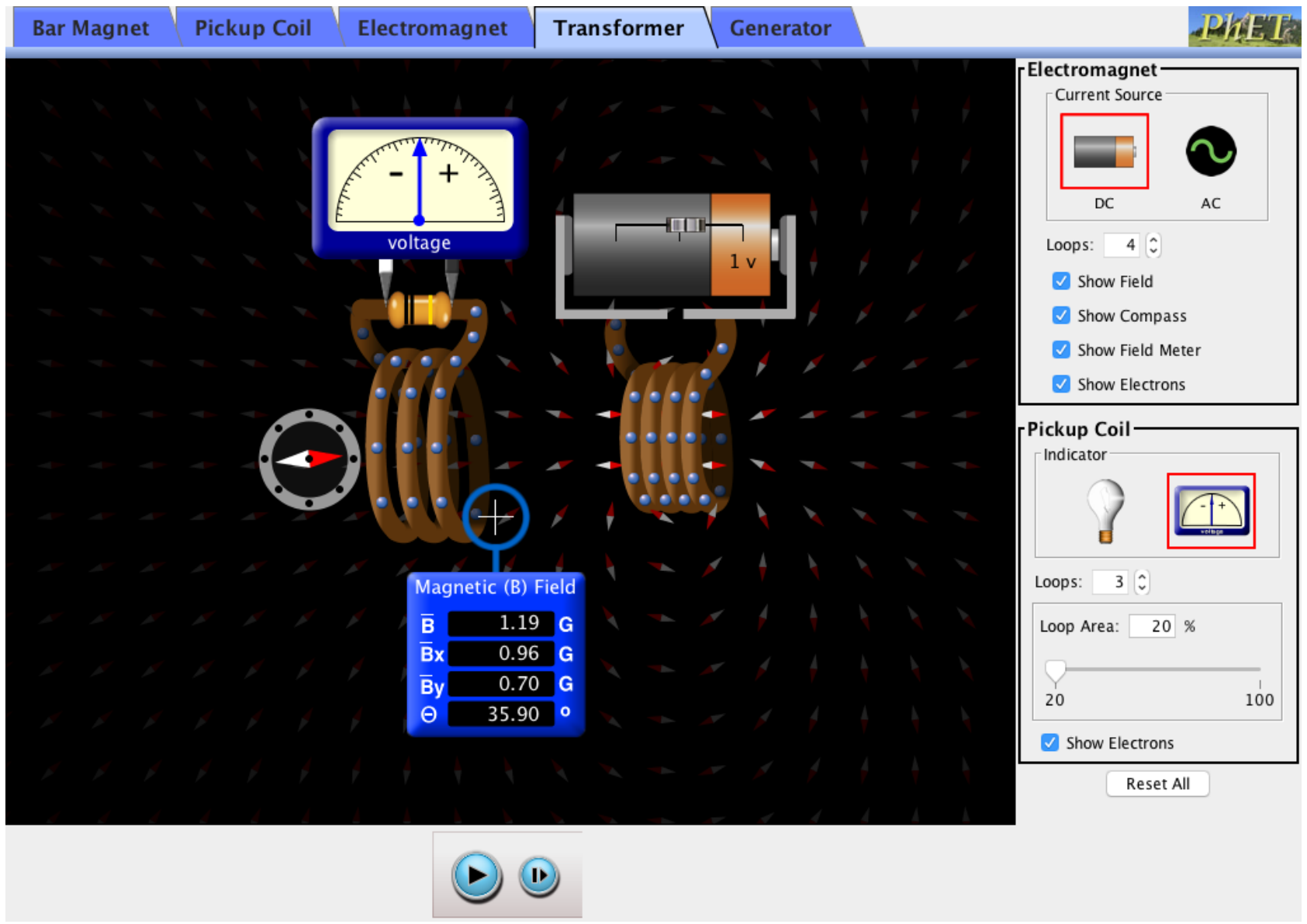

Figure 5 Simulation of the electromagnetic induction (Picture credit: https://phet.colorado.edu/)

*E. Hydro power production*

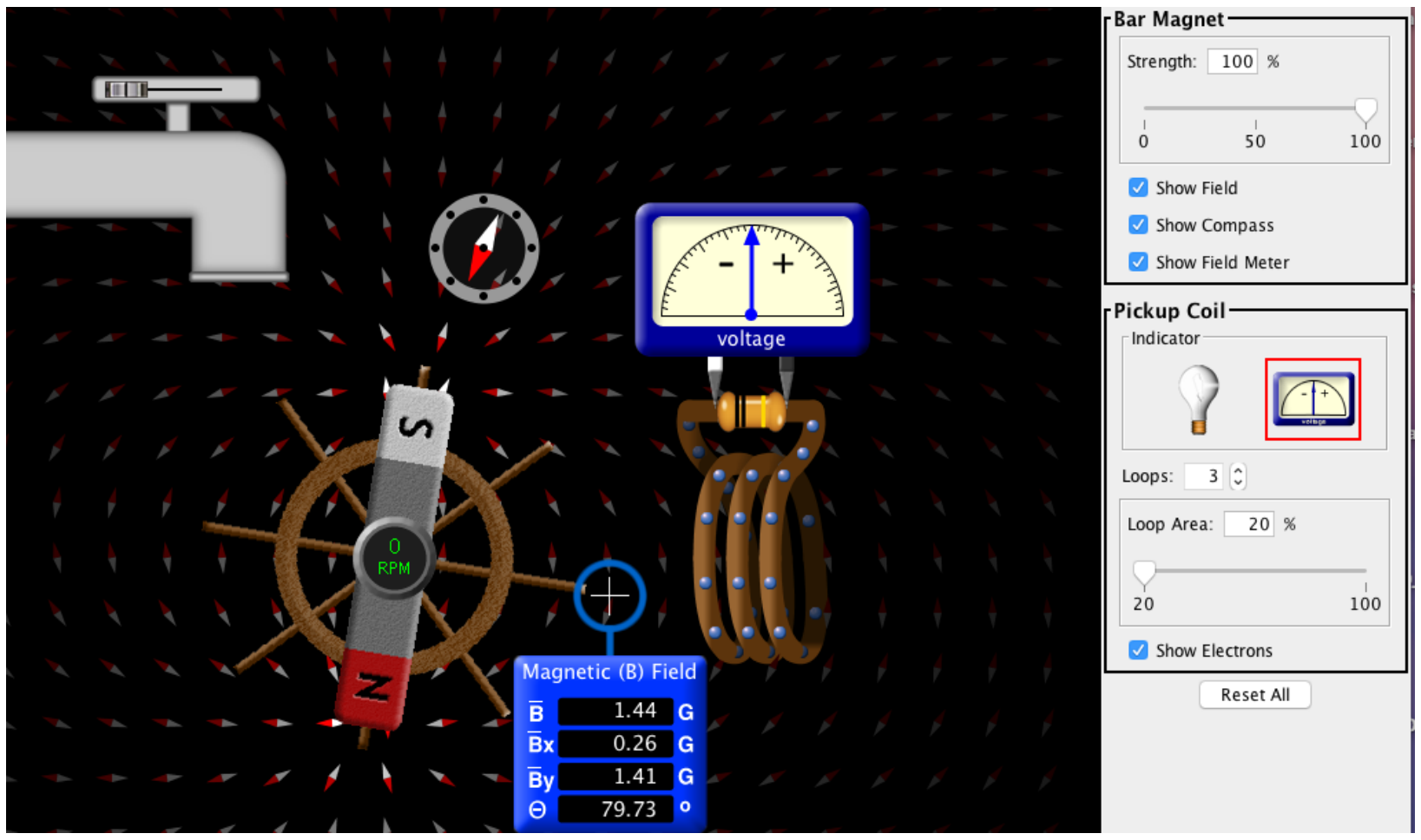

Figure 6 Simulation of hydro power plant (Picture credit: https://phet.colorado.edu/)

- Use the last option (transformer) in the simulation as shown in the figure.
- Set the bar magnet strength into 100%.
- Set the loop area of the secondary coil to 20%.
- Switch on the water flow slowly in the RPM (rotation per minute) of the bar magenta into 5.0.
- Observe the maximum deflection of the galvanometer attached to secondary coli and record it.
- Repeat the last two of the above procedures by increasing RPM by 10.0 at a time (from zero to 100 RPM).
- Make a graph of induced emp (galvanometer reading) vs RPM.
- Discuss the results in terms of Faraday and Lenz laws in table-5 on your report.
- Also discuss the behavior of induced emf in the secondary coil as s function of RPM of the bar magnet.

- Also answer the following questions.
    - When the bar magnet rotates what is changing inside the secondary coil?
    - What is the maximum induced emf of secondary coil by changing the cross section area of secondary coil to 100%?
    - What will happen to induced emf when the area of the secondary coil is increasing?
- What form of the initial energy changes its form to produce electric energy in this experiment?
- Does this experiment give you an idea of alternating-current (AC)?
- Discuss why the main power supply to your home is AC (alternating current) form?

**PRE LAB QUESTIONS**

1) Describe magnetic flux?
2) Describe Faraday's law of electromagnetic induction?
3) Describe Lenz's law?
4) Describe physical parameters that can be used in electromagnetic induction?

**POST LAB QUESTIONS**

1) How does the direction of the induced current depend on which pole of the permanent magnet is inserted into the solenoid?
2) How does the magnitude of the induced current depend on the strength of the bar magnet?
3) If the secondary solenoid is in a circuit with resistance *R,* there will be an induced current, $I_2 = \frac{\varepsilon}{R}$. Discuss how the equations predict the direction of the current in the secondary coil depending on whether the current in the *primary* coil, $I1$, is increasing or decreasing.
4) Does this match your observations? (Hint: Consider the effect of the sign of $\frac{dI_1}{dt}$ . Is it increasing or decreasing?)
5) If an alternating current source is supplied to the primary coil and the switch is closed (let the circuit work contentiously), then what kind of observation are you expecting to see in the secondary coil?

## DATA ANALYSIS AND CALCULATIONS

A. *Graph of magnetic field lines around the bar magnet.*

- Investigate magnetic field lines of bar magnet by using a simulator and insert the pictures of magnetic field graphs around the bar magnet.

Table 1 Magnetic field map around bar magnet

| Strength of bar magnet is 25% | Strength of bar magnet is 50% |
|---|---|
| | |
| Strength of bar magnet is 75% | Strength of bar magnet is 100% |
| | |

B. *Using stronger bar-magnet to induce current in a Solenoid*

Table 2 Stronger bar magnet moves into the Solenoid

| | | Tral-1 | Trial-2 | Tral-3 | |
|---|---|---|---|---|---|
| North pole into the Solenoid | Direction of current in Galvanometer | Max Deflection in Galvanometer [ ] | Max Deflection in Galvanometer [ ] | Max Deflection in Galvanometer [ ] | Average Max Deflection in Galvanometer [ ] |
| Moves quickly inward | | | | | |
| Stationary inside | | | | | |
| Moves quickly outward | | | | | |
| Moves slowly inward | | | | | |
| Moves slowly outward | | | | | |

Table 3 Stronger bar magnet moves into the Solenoid

| | | Tral-1 | Trial-2 | Tral-3 | |
|---|---|---|---|---|---|
| South pole into the Solenoid | Direction of current in Galvanometer | Max Deflection in Galvanometer [ ] | Max Deflection in Galvanometer [ ] | Max Deflection in Galvanometer [ ] | Average Max Deflection in Galvanometer [ ] |
| Moves quickly inward | | | | | |
| Stationary inside | | | | | |
| Moves quickly outward | | | | | |
| Moves slowly inward | | | | | |
| Moves slowly outward | | | | | |

C. *Induced magnetic field due to current in the solenoid.*

Table 4 Magnetic field due to current in solenoid

| *Positive* Battery voltage [ ] | Magnetic field direction | Magnetic field strength [ ] | *Negative* Battery voltage [ ] | Magnetic field direction | Magnetic field strength [ ] |
|---|---|---|---|---|---|
| | | | | | |
| | | | | | |
| | | | | | |
| | | | | | |
| | | | | | |
| | | | | | |
| | | | | | |

*D. Changing current in one solenoid to induce current in the secondary solenoid*

Table 5 Induce current in the secondary coil

| | Positive battery voltage in primary coil | | Negative battery voltage in primary coil | |
|---|---|---|---|---|
| Cross section area of secondary coil | Direction of galvanometer deflection | Max deflection of galvanometer [ ] | Direction of galvanometer deflection | Max deflection of galvanometer [ ] |
| 20% | | | | |
| 40% | | | | |
| 60% | | | | |
| 80% | | | | |
| 100% | | | | |

*E. Hydro power production*

Table 6 Induce emf and hydro power production

| Rotation per minute of the bar magnet RPM [ ] | Maximum positive deflection of galvanometer [ ] | Maximum negative deflection of galvanometer [ ] |
|---|---|---|
| | | |
| | | |
| | | |
| | | |
| | | |
| | | |
| | | |
| | | |
| | | |
| | | |

# EXPERIMENT 9 INTRODUCTION TO OSCILLOSCOPE

## OBJECTIVE

Use of Oscilloscope is investigated. Simple resistor circuit with different types of applied voltage signals is investigated. Voltage and frequency of the applied voltage are measured with Oscilloscope.

## THEORY AND PHYSICAL PRINCIPLE

Oscilloscopes can be used to visualize the oscillations of voltage or current which were originally developed by using cathode ray tubes. There are two types of oscilloscope, a) analogue oscilloscope (consists of cathode ray tube) and b) digital oscilloscope.

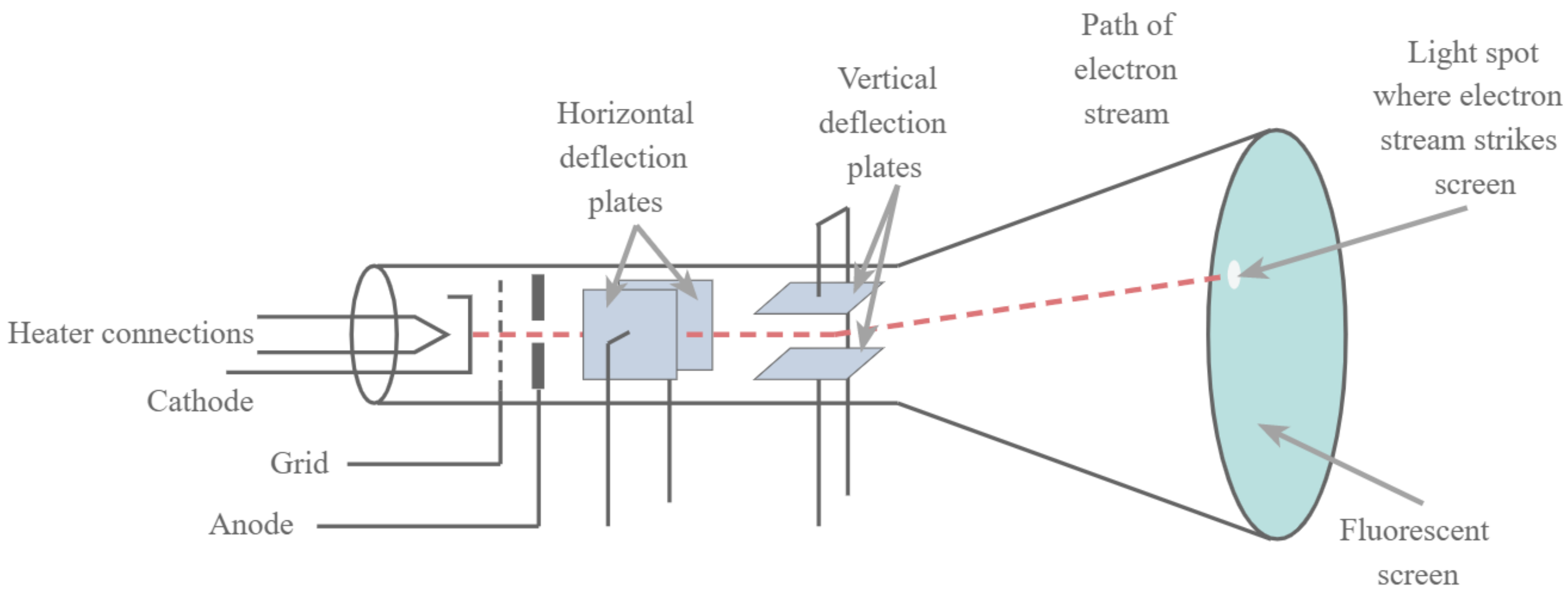

Figure 1 Cathode ray (analogue) oscilloscope (Picture credit: https://www.electronics-notes.com/)

Analogue oscilloscope (cathode ray oscilloscope or CRO) contains a cathode ray which uses electrostatic rather than magnetic deflection of an electron beam. This is important because an oscilloscope should be able to operate at extremely high frequencies. Cathode ray tubes were used in the first generation of televisions and those are different from the analogue oscilloscope. One of the primary differences between is the magnetic field deflection which was used only in the cathode ray tubes in televisions. [13-16]

Cathode ray tube in analogue oscilloscope display signals in both horizontal (X axis) and vertical (Y axis) and uses horizontal and vertical electrodes to deflect electron beams. Y axis displays the voltage value (in units of volts/division) and the X axis displays the oscillation of wave form (in units of time/division). [13-16]

Digital oscilloscope consists of an analogue-to-digital converter (ADC) which converts measured voltage into digital information. Therefore, digital oscilloscope does not contain cathode ray tube and size of the instrument is smaller and consists of many functionalities which is not available in analogue oscilloscopes. [13-16]

## APPARATUS AND PROCEDURE

- This experiment will be done by using a virtual electronic lab simulator. This simulation can be done on a web browser. Please click here to access the virtual simulations: https://www.multisim.com/
- This will ask you to create an account. You can get access to an online simulator after you create an account and log into the system.
- This is a very reputed company in the USA. You can check more info here, https://www.ni.com/en-us/shop/electronic-test-instrumentation/application-software-for-electronic-test-and-instrumentation-category/what-is-multisim.html
- If you like then you can download the software, but it has only Windows version.
- You can learn Multisim with a video tutorial. Check here: https://www.youtube.com/watch?v=xmJOzJb8SLU

- A very detail video lesson of virtual lab (data collection with simulator and data analysis with excel) can be found here: https://youtu.be/bHnsSsYAdVE

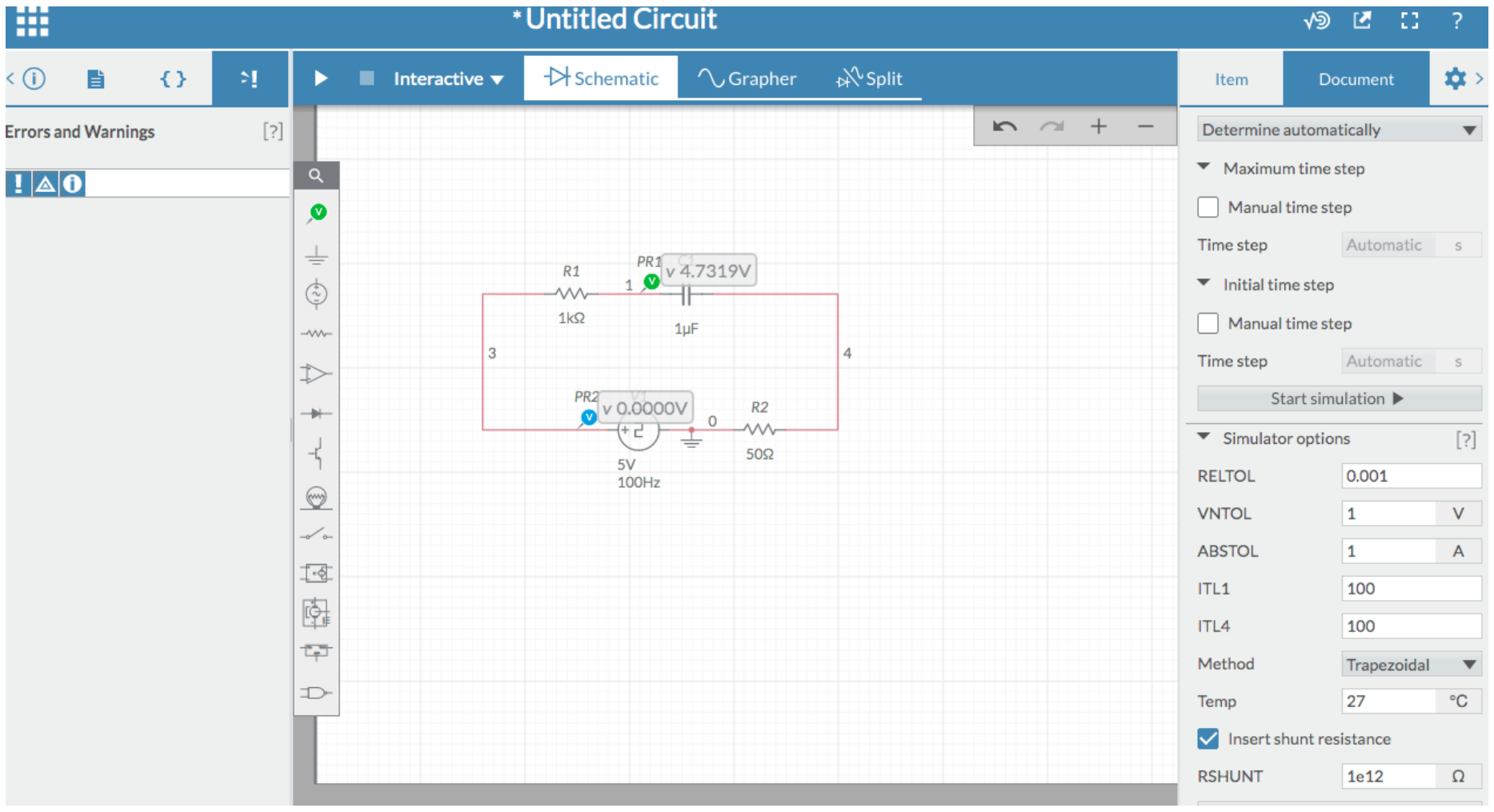

Figure 2 Multisim live online simulator (Picture credit: https://www.multisim.com/)

A. *Introduction to oscilloscope*
   - To learn more about the oscilloscope and its basic functionalities please watch the following video and include a summary of the video lesson to your report. Your report should include all the basic operations with details. Here is the link to short video: https://www.youtube.com/watch?v=dBVWv7enDsU

B. Use the virtual oscilloscope on the simulation website (Multisim) to measure different types of waves.
   - There is no oscilloscope icon in the virtual Multisim.
   - When the voltage sensor is attached to the circuit and simulates the circuit, in the graphing you will see the window exactly similar to the oscilloscope.
   - You have to set up a simple circuit. As shown in the figure-2.

- When you simulate then you will get the resultant graphs in figure-3. It is the oscilloscope itself and all the functions can be found on the right side corner. So, you can do all the adjustments as in the real oscilloscope.
- Record the applied voltage and frequency of the power supply (it is the signal generator in real experiment) in table-1.
- Measure the frequency and voltage of the signal generator through an oscilloscope.
- Select the trigger setting to “auto”.
- When measuring the frequency in the Multisim-oscilloscope, you may have to manually adjust the time (time-div) in the oscilloscope (x-axis).
- One thing to remember is that a Multisim-oscilloscope simulator does not reflect the correct units on x-axis. You have to measure the number of divisions per unit cycle in the window and then multiply it by the units of x-axis (time/div) with the correct unit.

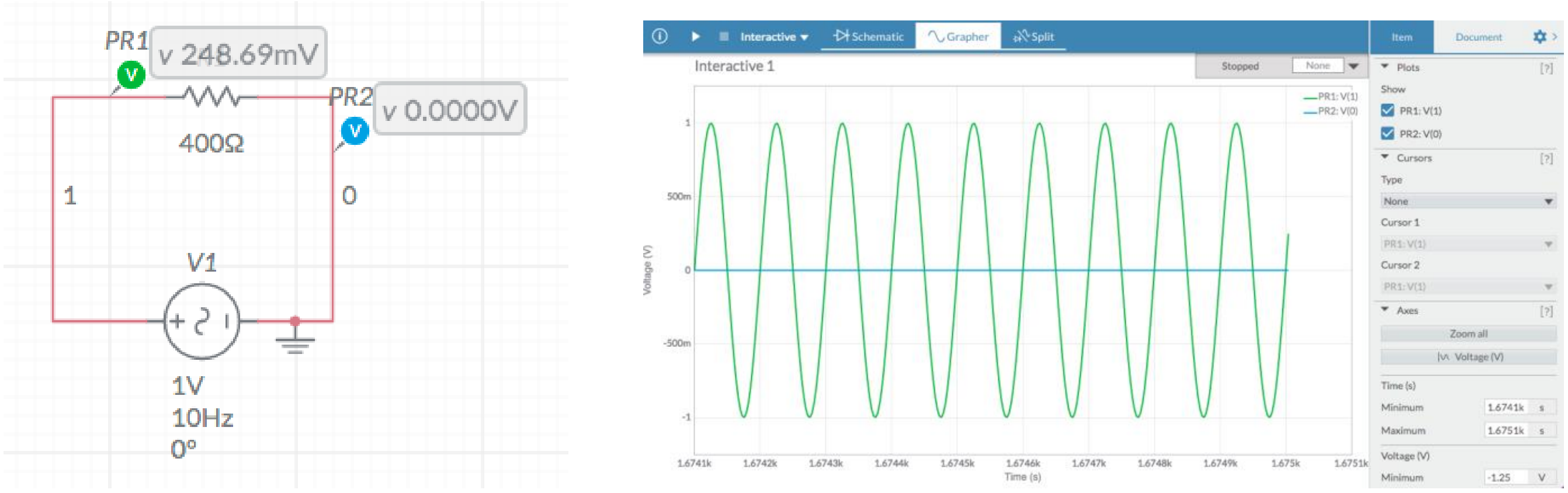

(a) (b)

Figure 3 (a) Virtual electronic simulation and (b) virtual oscilloscope with multisim-simulator (Picture credit: https://www.multisim.com/)

*C. Observing two of sine wave coupling on Oscilloscope (Lissajous Patterns)*

- When two sine waves are coupled in xy-mode, it will create a very interesting pattern in 3 dimensions. These patterns are called Lissajous figures.
- These cannot be done with a virtual oscilloscope, but it can be learned with a short video. https://www.youtube.com/watch?v=t6nGiBzGLD8
- Identify at least three different types of Lissajous figures and complete the following table with the information that you have learned with the video.

## PRE LAB QUESTIONS

1) Describe the difference between analogue and digital oscilloscopes?
2) What are the basic operations of an oscilloscope?
3) What does the Y axis represent in the oscilloscope screen?
4) What does the X axis represent in the oscilloscope screen?
5) What does the triggering mean?

## DATA ANALYSIS AND CALCULATION

A. *Introduction to oscilloscope*

Table 1 Details of functionalities of physical analogue oscilloscope

| Name of the switch | Describe the functionality |
|---|---|
| | |
| | |
| | |
| | |
| | |
| | |
| | |

B. Use the virtual oscilloscope on the simulation website (Multisim) to measure different types of waves.

Table 2 Sin wave analysis with Multisim-Oscilloscope

| Signal Generator Sine Wave | | Oscilloscope | | PE of observed and applied frequency | PE of observed and applied voltage |
|---|---|---|---|---|---|
| Frequency [ Hz ] | Voltage $V_P$ [ ] | Frequency [ ] | Voltage $V_P$ [ ] | % | % |
| 15 | 25 | | | | |
| 105 | 22 | | | | |
| 1005 | 18 | | | | |
| 10525 | 15 | | | | |
| 100505 | 12 | | | | |

- *Include a screenshot of the circuit with simulated sine waves.*

Table 3 Square wave analysis by Multisim-Oscilloscope

| Signal Generator Square Wave | | Oscilloscope | | PE of observed and applied frequency | PE of observed and applied voltage |
|---|---|---|---|---|---|
| Frequency [ Hz ] | Voltage $V_P$ [ ] | Frequency [ ] | Voltage $V_P$ [ ] | % | % |
| 15 | 25 | | | | |
| 105 | 22 | | | | |
| 1005 | 18 | | | | |
| 10525 | 15 | | | | |
| 100505 | 12 | | | | |

- *Include a screenshot of the circuit with simulated square waves.*

C. *Serial and Parallel Resistor Circuits with Multisim-Oscilloscope*

- Build the following resistor circuits and measure voltage across each of the resistors with the oscilloscope.
- *Include a screenshot of the circuit with simulated sine waves.*

Table 4 Circuit voltage analysis by Multisim-Oscilloscope

| Components | Measured voltage and current across each resistor with Oscilloscope | Calculated resistance of resistor (by using Ohm's law) | PD between measures and calculated resistance |
|---|---|---|---|
| Serially connected 3 resistors $R_1 = 15.0\Omega$ $R_2 = 45.0\Omega$ $R_3 = 65.0\Omega$ | | | |
| | | | |
| | | | |
| $R_1$ and $R_2$ serially and $R_3$ parallel to above $R_1 = 15.0\Omega$ $R_1 = 45.0\Omega$ $R_1 = 65.0\Omega$ | | | |
| | | | |
| | | | |
| $R_2$ and $R_3$ serially and $R_1$ parallel $R_1 = 15.0\Omega$ $R_1 = 45.0\Omega$ $R_1 = 65.0\Omega$ | | | |
| | | | |
| | | | |

D. *Observing two of sine wave coupling on Oscilloscope (Lissajous Patterns)*

- When two sine waves are coupled in xy-mode, it will create a very interesting pattern in 3 dimensions. These patterns are called Lissajous figures.
- These cannot be done with a virtual oscilloscope but it can be learned with a short video. https://www.youtube.com/watch?v=t6nGiBzGLD8
- Identify at least three different types of Lissajous figures and complete the following table with the information that you have learned with the video.

Table 5 Analysis of Lissajous figures

| Case Number | Frequency of Ch-1 [ ] | Frequency of Ch-2 [ ] | Draw a picture of the Lissajous pattern |
|---|---|---|---|
| 1 | | | |
| 2 | | | |
| 3 | | | |

# EXPERIMENT 10 RC CIRCUIT WITH OSCILLOSCOPE

## OBJECTIVE

A circuit with a capacitor and a resistor is investigated by using an oscilloscope. Capacitor charging and discharging graphs are analyzed and time constant is measured by using an oscilloscope. Circuit with different combinations of capacitors is investigated.

## THEORY AND PHYSICAL PRINCIPLES

When a capacitor connects with DC voltage (battery or square wave function) the capacitor builds a potential across it up to the maximum of the applied voltage which is called the charging circuit. If a fully charged capacitor connected with a resistor without a voltage in the circuit then the capacitor voltage produces a current through the circuit which discharges the capacitor and is called the discharging circuit.

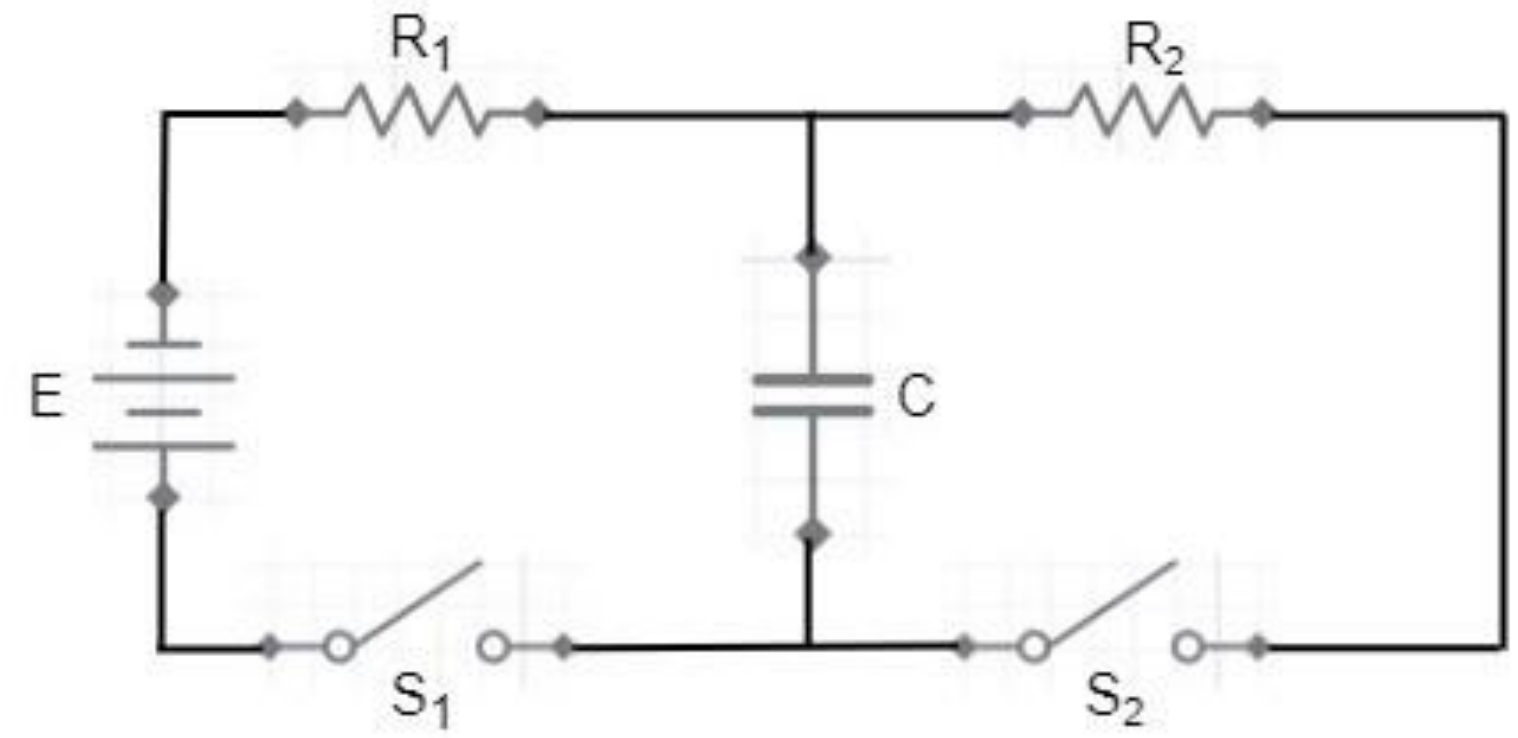

Figure 1 Charging and discharging capacitor circuits

In figure 1, capacitor charging when switch $S_1$ closes and switch $S_2$ open. When apply Kirchhoff's loop rule to charging circuit, (E is applied voltage, $V_c$ voltage drop across capacitor and IR is voltage drop across resistor)

$$E - V_c - IR = 0 \tag{1}$$

$$E - \frac{q}{C} - R\frac{dq}{dt} = 0 \tag{2}$$

$$\frac{dq}{dt} - \frac{q-EC}{RC} = 0 \tag{3}$$

By solving equation (3), current through the circuit can be found as follows.

$$q(t) = q_0\left(1 - e^{-t/\tau}\right) \tag{4}$$

$$q_0 = CE \text{ and } \tau = RC = \text{time constant} \rightarrow \text{ units equal to seconds } (s)$$

By dividing both sides by the capacitance C, equation (4) can be converted into an equation of voltage across the charging capacitor.

$$V_c(t) = V_0\left(1 - e^{-t/\tau}\right) \tag{5}$$

$$V_0 = E$$

Maximum potential across the capacitor when it is fully charged must be equal to the applied potential of E.

Time constant can be measured by using voltage vs time graph charging capacitor.

When $t = \tau, equation\ (4) \rightarrow V_c(t) = 0.63V_0$.

By derivative with respect to time equation (4) current in the charging circuit can be found. τ

$$I(t) = I_0 e^{-t/\tau} \qquad (6)$$

$$I_0 = \frac{E}{R}$$

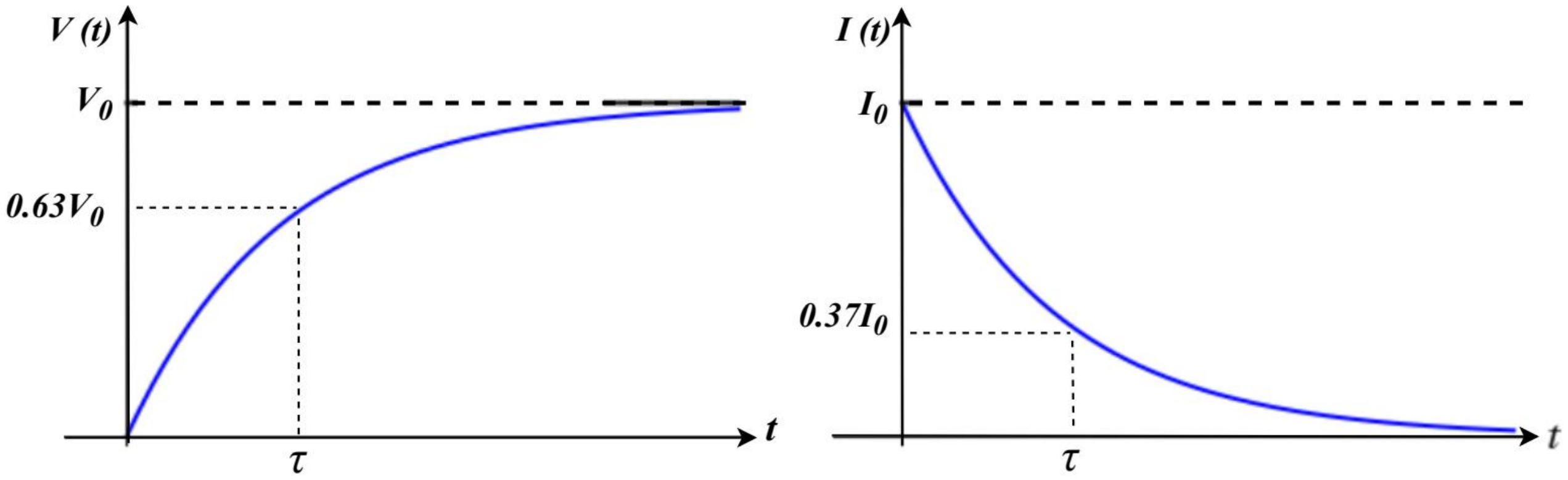

Figure 2 Voltage and current behavior of charging circuit

Figure 2 shows the voltage and current behavior across the charging capacitor. When applying time equal to one time constant ($t$ = τ) to equations (4) and (5) it can be found that the capacitor charges about 67% of total voltage within one time constant and current drops to about 33% of current through the capacitor.

Now consider discharging the capacitor circuit and applying Kirchhoff's loop rule.

$$V_c + IR = 0 \qquad (7)$$

$$\frac{q}{C} + R\frac{dq}{dt} = 0 \qquad (8)$$

$$\frac{dq}{dt} + \frac{q}{RC} = 0 \qquad (9)$$

By solving equation (3), current through the circuit can be found as follows.

$$q(t) = q_0 e^{-t/\tau} \qquad (10)$$

By dividing both sides by the capacitance C, equation (9) can be converted into an equation of voltage across the charging capacitor.

$$V_c(t) = V_0 e^{-t/\tau} \qquad (11)$$

Time constant can be measured by using voltage vs time graph discharging capacitor. When $t = \tau, equation\ (10) \rightarrow V_c(t) = 0.37V_0$.

Maximum potential across the capacitor when it is fully charged must be equal to the applied potential of E.
By derivative with respect to time equation (9) current in the charging circuit can be found. τ

$$I(t) = -I_0 e^{-t/\tau} \qquad (12)$$

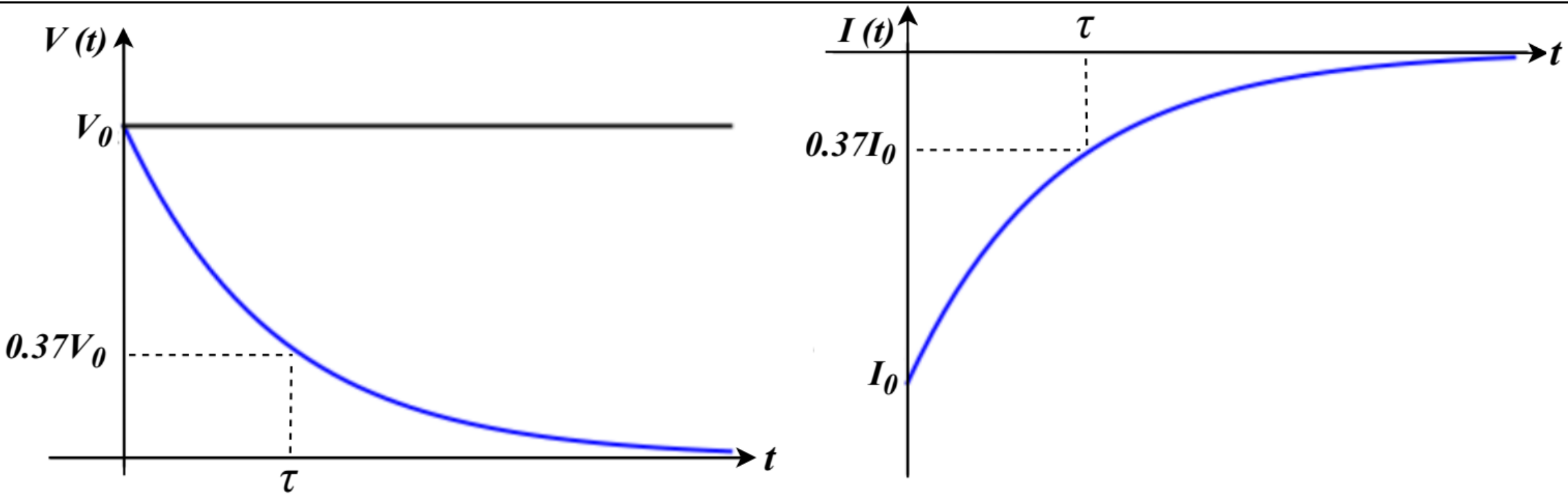

Figure 3 Voltage and current behavior of discharging circuit

## APPARATUS AND PROCEDURE

- This experiment will be done by using a virtual electronic lab simulator. This simulation can be done on a web browser. Please click here to access the virtual simulations: https://www.multisim.com/
- This will ask you to create an account. You can get access to an online simulator after you create an account and log into the system.
- You can learn Multisim with a video tutorial. Check here: https://www.youtube.com/watch?v=xmJOzJb8SLU
- After you get into the live online Multisim, it should look like following:

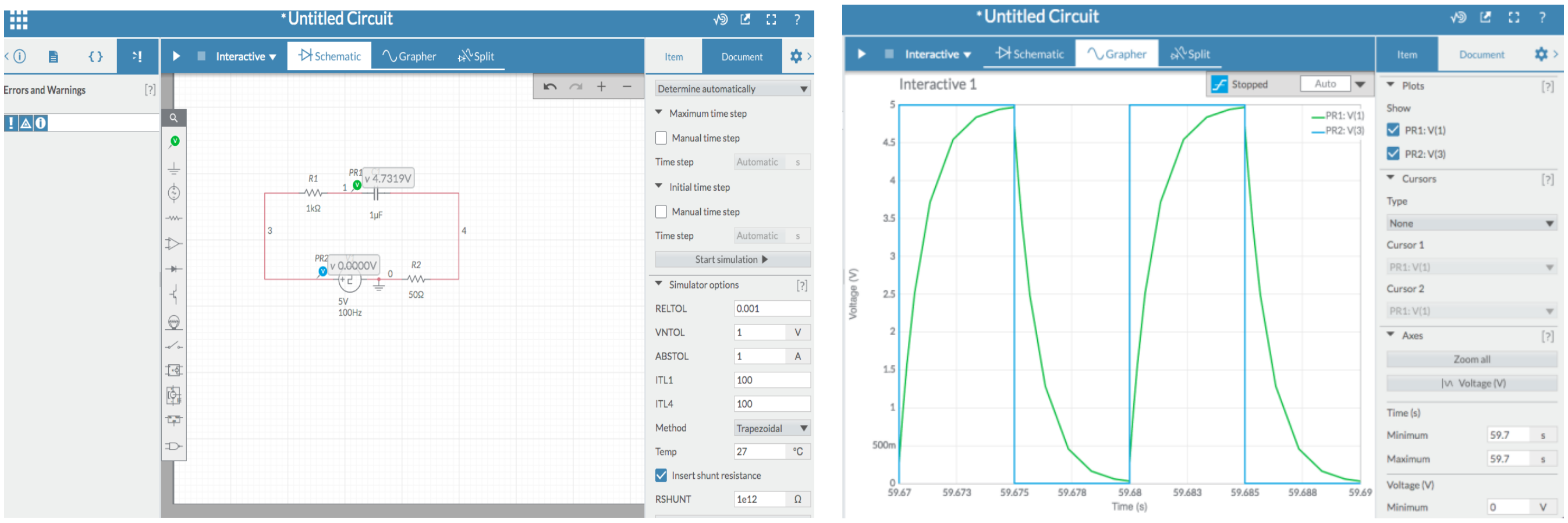

(a) (b)

Figure 4 Multisim live online simulator, (a) circuit maker and (b) Oscilloscope simulator (Picture credit: https://www.multisim.com/)

*RC charging and discharging circuit*

- A very detail video lesson of virtual lab (data collection with simulator and data analysis with excel) can be found here: https://youtu.be/BRDak7ZoP2M
- Build each of the RC circuits in the following table.
- Add a picture of the circuit from Multisim to table-1.
- Adjust all the values in the simulation with given values of resistors and capacitors.
- Add voltage-measuring probes to measure input and output voltages across the capacitor.
- Simulate the circuit and measure the time constant by Multisim and record in table-2.

## DATA ANALYSIS AND CALCULATIONS

Table 1 RC circuits diagrams from Multisim simulator

| Values of resistor and capacitor | Draw circuit diagram from Multisim |
|---|---|
| Circuit-1<br><br>R1 = 100.0 kΩ<br>C = 10.0 nF | |
| Circuit-2<br><br>serial R1 + R2<br>R1 = 100.0 kΩ<br>R1 = 50.0 kΩ<br><br>C = 10.0 nF | |
| Circuit-3<br><br>parallel R1 and R2<br>R1 = 100.0 kΩ<br>R1 = 50.0 kΩ<br>C = 10.0 nF | |
| Circuit-4<br><br>R1 = 10 kΩ<br><br>serial C1 and C2<br>C1 = 10.0 nF<br>C2 = 15.0 nF | |
| Circuit-5<br><br>R1 = 10 kΩ<br><br>parallel C1 and C2<br>C1 = 10.0 nF<br>C2 = 15.0 nF | |

Table 2 Time constant analysis of RC circuits

| Values of resistor and capacitor | Calculate (analytically) $\tau_{cal}$ [ ] | Estimate (charging simulation) $\tau_{est_1}$ [ ] | Estimate (discharging Simulation) $\tau_{est_2}$ [ ] | Estimate average $\tau_{avg}$ [ ] | PD between $\tau_{cal}$ and $\tau_{avg}$ |
|---|---|---|---|---|---|
| Circuit-1<br>R1 = 100.0 kΩ<br>C = 10.0 nF | | | | | |
| Circuit-2<br>serial R1 + R2<br>R1 = 100.0 kΩ<br>R1 = 50.0 kΩ<br>C = 10.0 nF | | | | | |
| Circuit-3<br>parallel R1 and R2<br>R1 = 100.0 kΩ<br>R1 = 50.0 kΩ<br>C = 10.0 nF | | | | | |
| Circuit-4<br>R1 = 10 kΩ<br>serial C1 and C2<br>C1 = 10.0 nF<br>C2 = 15.0 nF | | | | | |
| Circuit-5<br>R1 = 10 kΩ<br>parallel C1 and C2<br>C1 = 10.0 nF<br>C2 = 15.0 nF | | | | | |

# EXPERIMENT 11 RLC CIRCUITS AND IMPEDANCE

## OBJECTIVE

Reactance of capacitor and inductor are investigated by using simple AC circuits of RC and RL. Also, impedance of the RLC circuit is investigated. Behavior of current as a function of frequency and maximum current through the RLC circuit are investigated.

## THEORY AND PHYSICAL PRINCIPLES

If direct-current (DC) applies to a resistor-capacitor (RC) circuit then the capacitor acts like a circuit breaker because when the capacitor fully charged it stops the current DC current passing through the RC circuit. When AC current applies to a resistor-inductor (RL) circuit then the inductor acts like a simple resistor. On the other hand, if alternating-current (AC) applies through a capacitor or inductor then the current pass through depends on frequency of the applied voltage which develops reactance to AC current.

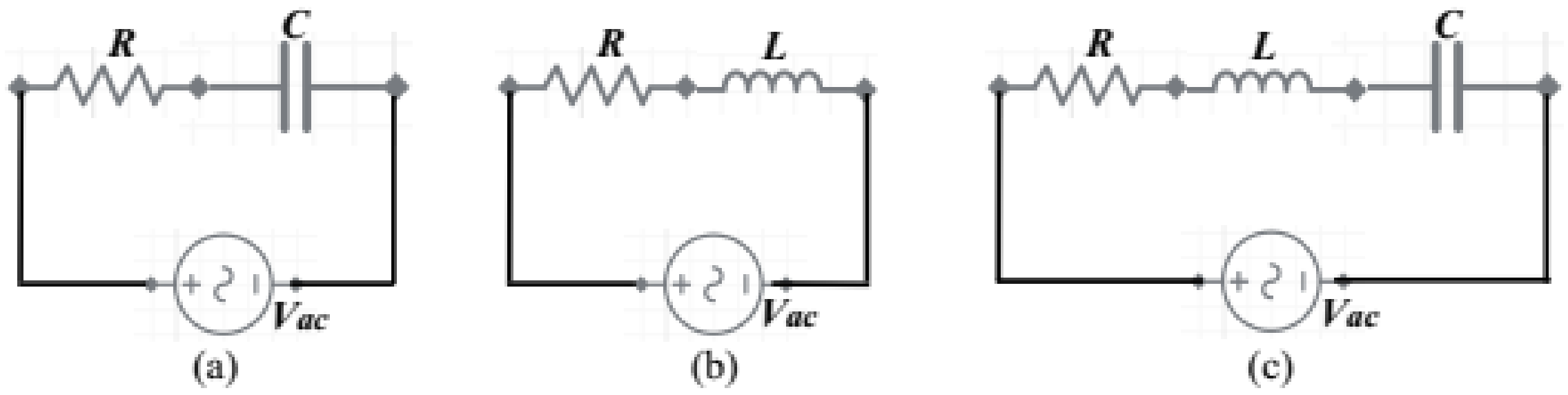

Figure 1 Circuit diagrams of (a) RC circuit, (b) RL circuit and (c) RLC circuit with applied AC voltage

When an AC voltage of frequency of $f$ applies to a capacitor of capacitance of $C$, reactance ($X_c$) of a capacitor can be written as follows.

$$X_C = \frac{1}{2\pi f C} \qquad (1)$$

When an AC voltage of frequency of $f$ applies to an inductor of inductance of $L$, reactance ($X_L$) of inductor can be written as follows.

$$X_L = 2\pi f L \qquad (2)$$

Voltage across each component can be written as,

$$V_R = IR \qquad (3)$$

$$V_C = IX_C = \frac{I}{2\pi f C} \qquad (4)$$

$$V_L = IX_L = I2\pi f L \qquad (5)$$

When an AC voltage and current applies into circuit then those may be in phase to each other therefore phasor diagram is used to explain the behavior and is used to find the resultant voltage across combinations of resistor, capacitor, and inductor. AC voltage and current through a resistor are in phase to each other. AC voltage is $90^0$ degrees delay (behind) of the AC current through a capacitor and AC voltage is $90^0$ degrees earlier (in front) through the inductor.

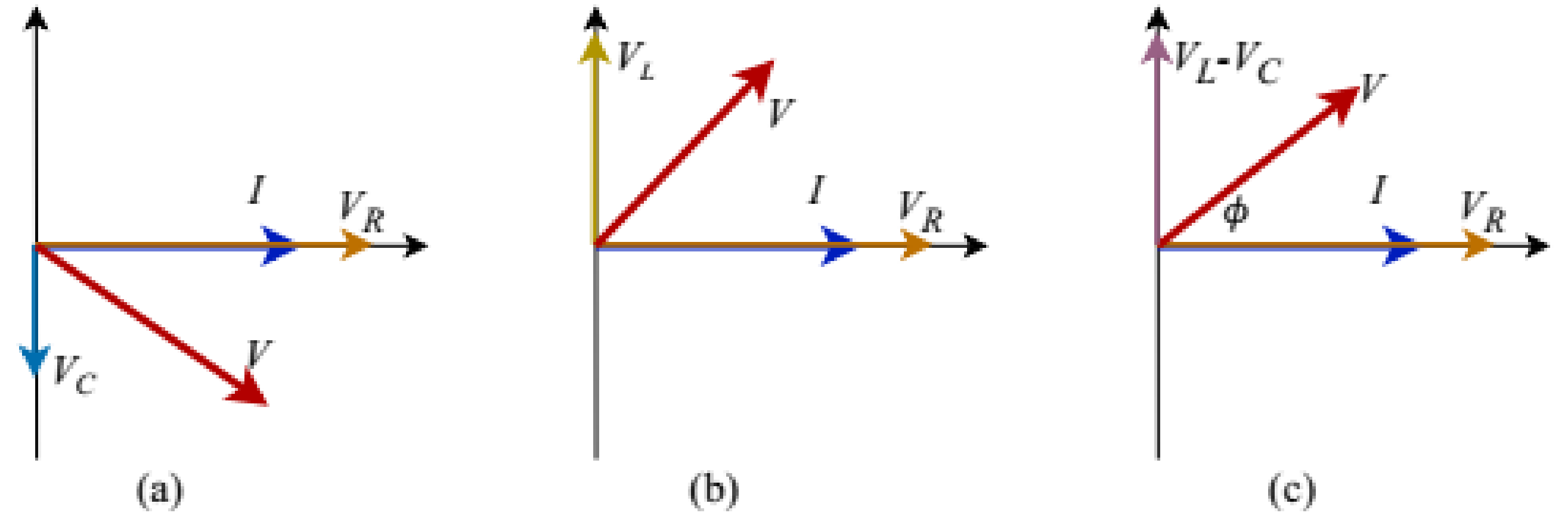

Figure 2 Phasor diagrams of (a) RC circuit, (b) RL circuit and (c) RLC circuit with applied AC voltage

Total voltage drop across the RLC series circuit can be found by using the phase diagram of figure 2(c).

$$V = \sqrt{V_R^2 + (V_L - V_C)^2} \tag{6}$$

If Ohm's law is applied to equivalent AC reactance (impedance=$Z$) for series RLC circuit,

$$V = IZ \tag{7}$$

By combining the above equations (3,4,5,6,7), it is possible to find an equation for impedance of series RLC circuit.

$$Z = \sqrt{R^2 + (X_L - X_C)^2}$$

$$Z = \sqrt{R^2 + \left(2\pi fL - \frac{1}{2\pi fC}\right)^2} \tag{8}$$

Current pass through RLC series circuit can be written as,

$$I = \frac{V}{Z} = \frac{V}{\sqrt{R^2 + \left(2\pi fL - \frac{1}{2\pi fC}\right)^2}} \tag{9}$$

Equation (9) shows that the current in the RLC series circuit is maximum when the denominator of the equation is minimum.

$$I_{max} \rightarrow when \sqrt{R^2 + \left(2\pi fL - \frac{1}{2\pi fC}\right)^2} \rightarrow minimum$$

$$I_{max} \rightarrow when\ 2\pi fL - \frac{1}{2\pi fC} = 0$$

$$f = \frac{1}{2\pi\sqrt{LC}} \tag{10}$$

When the current passes through the RLC series circuit maximum it is called the resonance and the resonance frequency of the circuit can be found by using equation (10).

## APPARATUS AND PROCEDURE

- This experiment will be done by using a virtual electronic lab simulator. This simulation can be done on a web browser. Please click here to access the virtual simulations: https://www.multisim.com/
- You can learn Multisim with a video tutorial. Check here: https://www.youtube.com/watch?v=xmJOzJb8SLU
- A very detail video lesson of virtual lab (data collection with simulator and data analysis with excel) can be found here: https://youtu.be/iK6VUg5HtPM

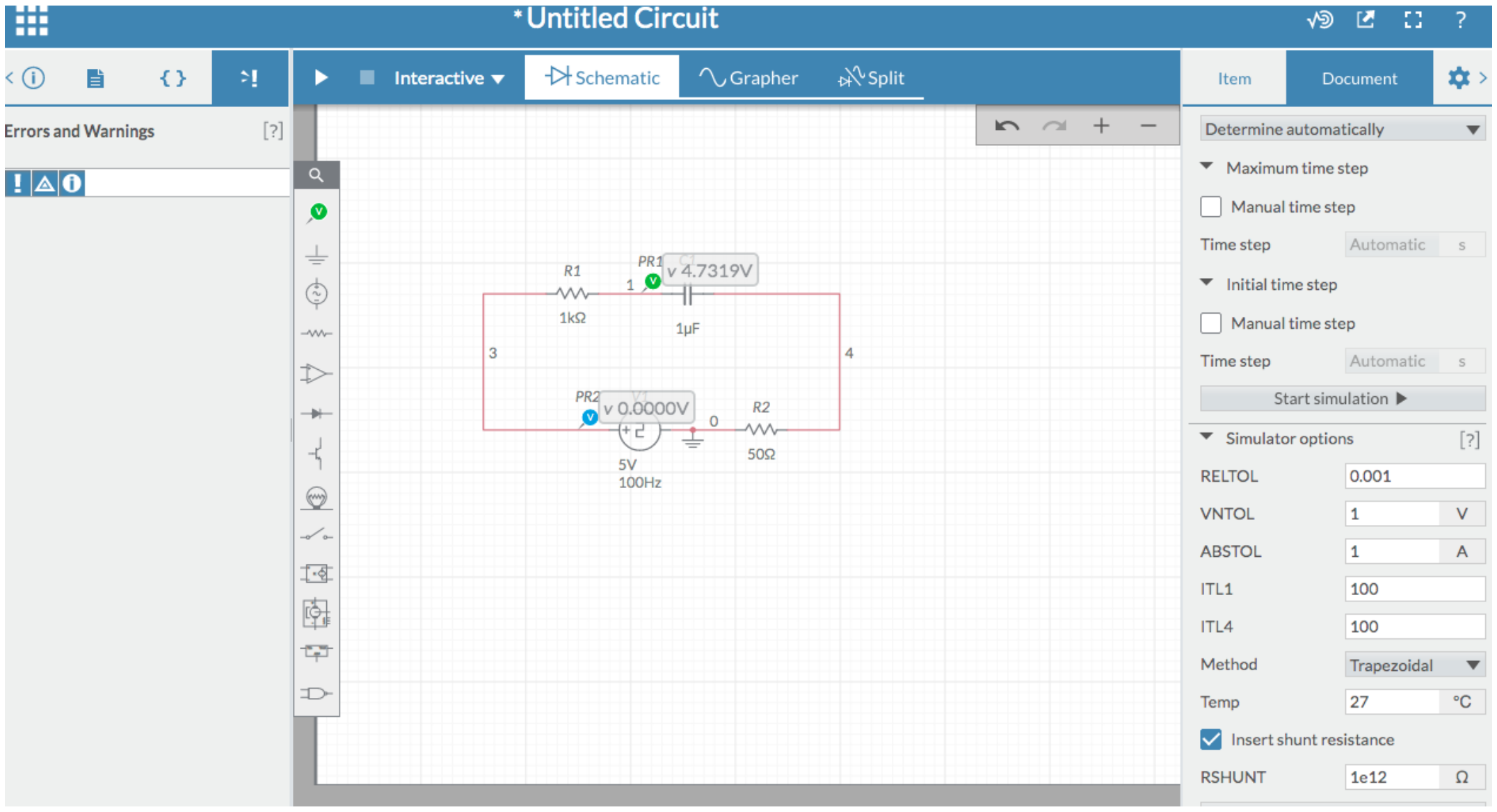

Figure 3 Multisim live online simulator (Picture credit: https://www.multisim.com/)

*A. RC Circuit (R=45 Ω, C=0.10 μF)*

- Build the circuit with a Multisim-simulator.
- Attached the picture of the circuit to your report.
- Find an algebraic equation for total impedance of the RC circuit.
- Simulate the circuit with given frequencies in table-1 and measure the current and voltage across the capacitor.
- Record the current and voltage across the capacitor in tebal-1.
- Calculate the reactance of the capacitor by using applied frequency (Xc_cal) and measuring voltage and current (Xc_meas).
- Make a graph of reactance (Xc_measured) vs frequency (*f*).
- Do the fitting with y=a/x (inverse law behavior) and find the capacitance ($C_{graph}$) of the capacitor?
- Find the percent error of capacitance $C_{graph}$ with the actual value (C)?
- Make a graph of reactance (Xc_measured) vs 1/frequency (1/*f*).
- Do the fitting with y=ax (linear behavior) and find the capacitance ($C_{graph}$) of the capacitor?
- Find the percent error of capacitance $C_{graph}$ with the actual value (C)?

*B. RL Circuits (R=45 Ω, L=15 mH)*

- Build the circuit with a Multisim-simulator.
- Attached the picture of the circuit to your report.
- Find an algebraic equation for total impedance of the RC circuit.
- Simulate the circuit with given frequencies in table-2 and measure the current and voltage across the capacitor.
- Record the current and voltage across the capacitor in tebal-1.
- Calculate the reactance of the inductor by using applied frequency ($X_L$_cal) and measures voltage and current ($X_L$_meas).
- Make a graph of reactance ($X_{L_meas}$) vs frequency (*f*).
- Do the fitting with a linear equation and find the inductance ($L_{graph}$) of the inductor?
- Find the percent error of inductance with the actual value (L)?

*C. RLC Series Circuit (R=45 Ω, C=0.10 μF, L=15 mH)*

- Build the circuit with a Multisim-simulator.
- Attached the picture of the circuit to your report.
- Find an algebraic equation for total impedance of the RLC circuit.
- Simulate the circuit with given frequencies in table-3 and measure the current and voltage across the capacitor.
- Calculate the reactance of the circuit by using applied frequency.
- Make a graph of current (***I***) vs frequency (***f***) and explain the behavior of the graph?
- Find the resonance frequency ($f_{res-graph}$) by using the maximum value of the graph?
- Find the percent error between $f_{res-graph}$ and the calculated value ($f_{res_cal}$)?

## PRE LAB QUESTIONS

1) Describe the reactance of a capacitor when it is connected to AC voltage?
2) Describe the reactance of an inductor when it is connected to AC voltage?
3) Describe the effective reactance (impedance) of a resistor-capacitor (RC) serial circuit when connected to AC voltage?
4) Describe the effective reactance (impedance) of a resistor-inductor (RL) serial circuit when connected to AC voltage?
5) Describe the effective reactance (impedance) of a resistor-inductor-capacitor (RLC) serial circuit when connected to AC voltage?
6) Describe the resonance of series RLC circuit when connected to AC voltage?

# DATA ANALYSIS AND CALCULATIONS

*A. RC Circuits*

Table 1 Current and Frequency measurement of RC circuit

| Frequency (f) [ Hz ] | Current (Ic) [ ] | Voltage (Vc) [ ] | $\frac{1}{f}$ (s) [ ] | Reactance (Xc_cal) [ ] | Reactance (Xc_meas) [ ] | PD between Xc_cal and Xc_meas |
|---|---|---|---|---|---|---|
| 1000 | | | | | | |
| 2000 | | | | | | |
| 3000 | | | | | | |
| 4000 | | | | | | |
| 5000 | | | | | | |
| 6000 | | | | | | |
| 7000 | | | | | | |
| 8000 | | | | | | |
| 9000 | | | | | | |
| 10000 | | | | | | |

- Calculate the reactance of the capacitor by using applied frequency (Xc_cal) and measuring voltage and current (Xc_meas).
- Make a graph of current (***I***) vs frequency (***f***).
- Fit the data with the analytical equation given in the theory section and explain the behavior of the graph?
- Make a graph of reactance (Xc_measured) vs frequency (***f***).
- Do the fitting and find the capacitance ($C_{graph}$) of the capacitor?
- Find the percent error of capacitance $C_{graph}$ with the actual value (C)?

*B. RL Circuit*

Table 2 Current and Frequency measurement of RL circuit

| Frequency (f) [ Hz ] | Current ($I_L$) [ ] | Voltage ($V_L$) [ ] | Reactance calculated ($X_{L_calc}$) [ ] | Reactance measured ($X_{L_meas}$) [ ] | PD between $X_{L_calc}$ and $X_{L_meas}$ |
|---|---|---|---|---|---|
| 1000 | | | | | |
| 2000 | | | | | |
| 3000 | | | | | |
| 4000 | | | | | |
| 5000 | | | | | |
| 6000 | | | | | |
| 7000 | | | | | |
| 8000 | | | | | |
| 9000 | | | | | |
| 10000 | | | | | |

- Calculate the reactance of the inductor by using applied frequency ($X_L$_cal) and measures voltage and current ($X_L$_meas).
- Make a graph of current (***I***) vs frequency (***f***).
- Fit the data with the analytical equation given in the theory section and explain the behavior of the graph?
- Make a graph of reactance ($X_{L_meas}$) vs frequency (***f***).
- Do the fitting and find the inductance ($L_{graph}$) of the inductor?
- Find the percent error of inductance with the actual value (L)?

*C. RLC Circuit*

- Calculate the impedance of the circuit by using applied frequency.
- Find the impedance of the circuit by using current and voltage of the circuit.
- Make a graph of current (***I***) vs frequency (***f***) and explain the behavior of the graph?
- Find the resonance frequency ($f_{res\text{-}graph}$) by using the maximum value of the graph?
- Find the percent error between $f_{res\text{-}graph}$ and the calculated value ($f_{res_cal}$)?

Table 3 Current and Frequency measurement of RLC circuit

| Frequency (f) [ Hz ] | Current, *I*_meas [ A ] | Impedance *Z*_cal [ ] | Impedance *Z*_maes [ ] | PD between Z_meas and Z_cal [ % ] |
|---|---|---|---|---|
| 500 | | | | |
| 1000 | | | | |
| 1500 | | | | |
| 2000 | | | | |
| 2500 | | | | |
| 3000 | | | | |
| 3200 | | | | |
| 3500 | | | | |
| 3700 | | | | |
| 3800 | | | | |
| 3900 | | | | |
| 3950 | | | | |
| 4000 | | | | |
| 4050 | | | | |
| 4100 | | | | |
| 4150 | | | | |
| 4200 | | | | |
| 4500 | | | | |
| 4800 | | | | |
| 5000 | | | | |
| 5500 | | | | |
| 6000 | | | | |